\documentclass[twocolumn]{aastex701}

\usepackage{amsmath}
\usepackage{bm}

\begin{document}

\title{A Radon Transform Perspective on Exoplanet Transits}

\author[orcid=0000-0002-4093-8736]{Shotaro Tada}
\affiliation{Institute of Space and Astronautical Science, Japan Aerospace Exploration Agency, 3-1-1 Yoshinodai, Chuo-ku, Sagamihara, Kanagawa 252-5210, Japan}
\email[show]{tada@ir.isas.jaxa.jp}

\begin{abstract}
Transit light curves are usually analyzed under the assumption that the transiting planet has a circular sky-projected silhouette.
However, planetary rotation, tides, rings, or atmospheric inhomogeneities can produce non-circular silhouettes.
This raises the question of what information transit light curves can provide about the underlying two-dimensional attenuation map.
In this paper, we show that, in a simple and transparent limit, the time derivative of the transit light curve during ingress or egress can be interpreted as a Radon-transform measurement of the planetary attenuation map, with the projection direction set by the local normal to the stellar limb.
This viewpoint makes the information content of a single transit clear.
Ingress and egress provide at most two projection angles, so the data constrain the Fourier transform of the attenuation map only along at most two radial slices, leaving a large null space.
Physical constraints on the attenuation values and shape priors can reduce the range of viable solutions, but non-uniqueness generally remains.
We further examine how realistic effects modify this picture.
In particular, small stellar-limb curvature introduces weak sensitivity to transverse Fourier-space structure around the ideal slices, a sensitivity that is absent in the strict Radon-transform limit.
These results provide a framework for understanding what transit light curves can and cannot reveal about non-circular planetary silhouettes.
\end{abstract}

\keywords{\uat{Exoplanets}{498} --- \uat{Exoplanet structure}{495} --- \uat{Transit photometry}{1709}}

\section{Introduction}

Transit light curves are usually modeled under the assumption that the transiting planet is spherical, or equivalently that its sky-projected silhouette is circular.
This approximation has been highly successful for inferring the basic properties of exoplanetary systems \citep[e.g.,][]{2002ApJ...580L.171M, 2003ApJ...585.1038S, 2010exop.book...55W}.
However, as photometric precision and cadence improve, it becomes increasingly important to ask what additional information is encoded in ingress and egress when the projected planetary silhouette is not exactly circular.

Several effects can produce a non-circular silhouette.
Rapid rotation can cause oblateness, while tides can induce ellipsoidal or triaxial distortions \citep{Seager_2002,Barnes_2003,Leconte_2011,Correia_2014}.
For gaseous planets, the apparent shape can also reflect atmospheric structure.
Spatial variations in temperature, composition, clouds, or hazes can make the apparent silhouette wavelength dependent and azimuthally asymmetric \citep{Fortney_2010,Dobbs-Dixon_2012,Kempton_2017}.

Motivated by such effects, several transit models have been developed beyond the standard circular-occultor approximation.
For example, \texttt{catwoman} models asymmetric silhouettes using two semicircles with different radii, which is useful for morning--evening terminator asymmetries \citep{2021AJ....162..165E,Jones2022}.
\texttt{harmonica} represents the projected planetary limb as a Fourier series in polar angle \citep{Grant_2023}.
\texttt{squishyplanet} and \texttt{eclipsoid} have extended transit modeling to non-spherical and ellipsoidal planets \citep{squishyplanet,Dholakia_2025}.
Other forward-modeling approaches include \texttt{pyPplusS}, \texttt{JoJo}, and \texttt{Yuti} \citep{Rein_2019,Liu_2025,Bhowmick_2024}.
Using such models, non-circular effects have begun to be reported observationally, for example atmospheric asymmetries and tidal deformation \citep{2024Natur.632.1017E,Murphy_2025,Fu_2025,Barros_2022,Akinsanmi_2024}.

At the same time, several studies have shown that shape inference from transit light curves is intrinsically degenerate.
General non-parametric approaches reveal degeneracies in the inverse problem \citep{Sandford_2019}.
For oblate or ellipsoidal planets, changes in orbital and stellar parameters can mimic part of the signal of a non-circular silhouette \citep{Barnes_2003,Liu_2024,Dholakia_2025,Cassese_2026}.
The central issue is therefore not only how to build flexible forward models, but also what a one-dimensional light curve can and cannot reveal about a two-dimensional shadow.

In this paper, we show that, in a simple and physically transparent limit, ingress and egress are Radon-transform measurements of the planetary attenuation map.
This provides a compact framework for understanding the information content and degeneracies of transit light curves.

The basic intuition that ingress and egress encode spatial information is not new.
Related ideas have long appeared in eclipse mapping \citep{Majeau_2012,deWit_2012,Challener_2023} and in studies of non-circular or non-parametric transit shadows \citep{Arkhypov_2021,Sandford_2019,Nachmani_2022}.
In particular, \citet{2024MNRAS.528..596B} used the same straight-edge approximation adopted here to analyze the observable spatial modes and the spatial resolution attainable with eclipse mapping.
Our contribution is to identify explicitly the Radon-transform structure that emerges in this limit and to use it to clarify the limited angular coverage of a single transit, the resulting
continuous null space, and the associated shape–orbit
degeneracies.
We focus on a leading-order limit that is analytically transparent and useful for building intuition, and then examine how this picture is modified by realistic effects such as limb darkening, finite exposure time, stellar-limb curvature, and time-dependent planetary maps.

Our main focus is on transits, where the relevant quantity is the planetary attenuation map or projected silhouette.
The corresponding formulation for secondary eclipse is closely analogous, with the attenuation map replaced by a planetary brightness distribution.
We therefore keep the main text centered on transits and summarize the secondary-eclipse case in Appendix \ref{app:secondary_eclipse}.

This paper is organized as follows.
Section~\ref{sec:formulation} derives the Radon-transform limit for transit light curves.
Section~\ref{sec:nullspace} discusses the associated information content and null space in the Fourier plane.
Section~\ref{sec:deviations} examines departures from the idealized limit.
Section~\ref{sec:shape_inference} discusses the implications for shape inference from transits.

\section{Forward Model and the Radon-Transform Limit}
\label{sec:formulation}

In this section, we formulate the forward model for transit light curves and derive the Radon-transform limit under a set of simplifying approximations.
In this limit, the time derivative of the light curve during ingress or egress can be interpreted as a Radon-transform measurement of the planetary attenuation map, with the projection direction determined by the local normal to the stellar limb.
The corresponding expressions for secondary eclipses are closely analogous and are summarized in Appendix~\ref{app:secondary_eclipse}.

\subsection{Forward-Model Formulation}
\label{sec:forward_model}

\begin{figure}[tb!]
\epsscale{1.1}
\plotone{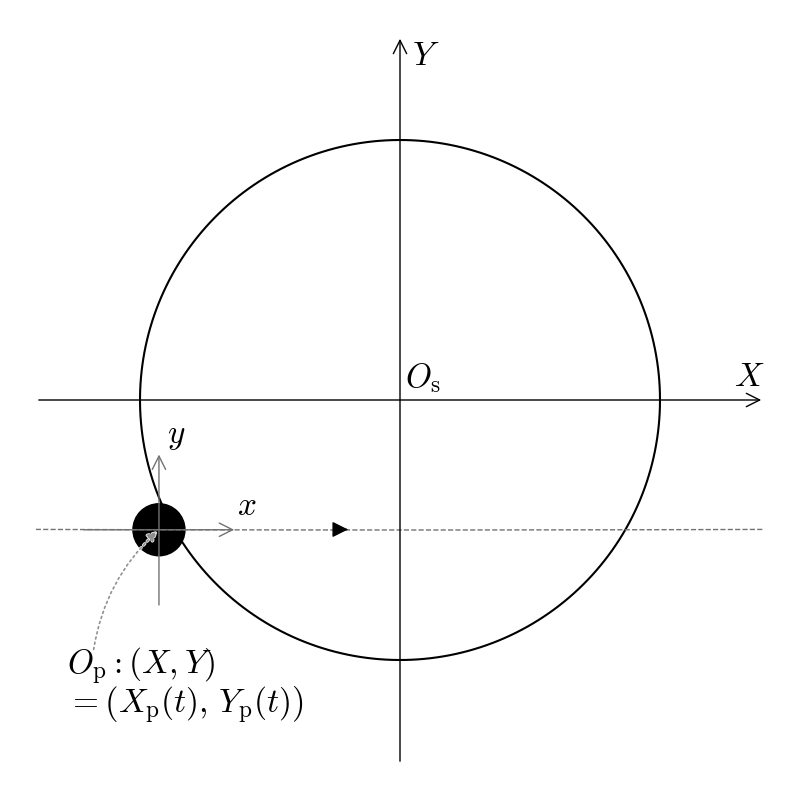}
\caption{Geometry and coordinate conventions.
The sky-plane coordinates $\bm{X}\equiv (X,Y)^{\mathrm T}$ are centered on the star, while $\bm{x}\equiv (x,y)^{\mathrm T}$ are centered on the planet.
The origins are denoted by $O_{\mathrm{s}}$ (stellar center) and $O_{\mathrm{p}}$ (planet center), respectively, with $O_{\mathrm{p}}$ located at $\left(X_{\mathrm{p}}(t),Y_{\mathrm{p}}(t)\right)$ in the $(X,Y)$ frame.
The large circle represents the projected stellar disk, and the black-filled shape represents the projected planetary silhouette.
The dashed line indicates the projected planetary trajectory, and the arrow indicates the direction of motion.
\label{fig:coordinate}}
\end{figure}

Let $\bm{x}\equiv (x,y)^{\mathrm T}$ be Cartesian coordinates on the sky plane whose origin is fixed to the planet's center of mass (Figure~\ref{fig:coordinate}).
In this planet-centered frame, we describe the planetary emission map by $I_{\mathrm{p}}(\bm{x})$.
To represent attenuation by the planetary body and atmosphere, we introduce the planetary transmittance map $\mathcal{T}_{\mathrm{p}}(\bm{x})$, where $\mathcal{T}_{\mathrm{p}}=1$ corresponds to fully transparent and $\mathcal{T}_{\mathrm{p}}=0$ to fully opaque.
We do not restrict the projected planetary silhouette to be circular.
Outside the projected planetary silhouette, we set $I_{\mathrm{p}}(\bm{x})=0$ and $\mathcal{T}_{\mathrm{p}}(\bm{x})=1$.
In principle, $I_{\mathrm{p}}$ and $\mathcal{T}_{\mathrm{p}}$ may be time dependent (e.g., because of rotation or evolving atmospheric patterns), but here we assume that they are effectively static over the duration of a single transit.
We discuss the consequences of relaxing this assumption in Section~\ref{sec:time_dependent_maps}.
We also define the complementary attenuation map
\begin{equation}
A_{\mathrm{p}}(\bm{x}) \equiv 1-\mathcal{T}_{\mathrm{p}}(\bm{x}).
\end{equation}

We next introduce a star-centered sky-plane coordinate system with position vector $\bm{X}\equiv (X,Y)^{\mathrm T}$, whose origin coincides with the stellar center (Figure~\ref{fig:coordinate}).
In this frame, the projected position of the planet's center of mass is denoted by
\begin{equation}
\bm{X}_{\mathrm{p}}(t)\equiv \left(X_{\mathrm{p}}(t),Y_{\mathrm{p}}(t)\right)^{\mathrm T}.
\end{equation}
Assuming that the two coordinate systems have no relative rotation on the sky plane, the coordinates are related by
\begin{equation}
\bm{X} = \bm{x} + \bm{X}_{\mathrm{p}}(t).
\end{equation}

In the star-centered frame, we specify the stellar intensity map $I_{\mathrm{s}}^{\star}(\bm{X})$ and set $I_{\mathrm{s}}^{\star}(\bm{X})=0$ outside the stellar disk.
We assume that the stellar map is static in the star-centered frame.
Expressed in planet-centered coordinates, it becomes time dependent as the star moves relative to the planet:
\begin{equation}
I_{\mathrm{s}}(\bm{x},t)
=
I_{\mathrm{s}}^{\star}\!\left(\bm{x}+\bm{X}_{\mathrm{p}}(t)\right).
\label{eq:Is_loc}
\end{equation}

In what follows, $I(\bm{x},t)$ denotes the sky-plane intensity reaching the observer at time $t$.
The observed flux is
\begin{equation}
F(t) \equiv \iint_{\mathbb{R}^2} I(\bm{x},t)\,dx\,dy,
\label{eq:F_def}
\end{equation}
and, for brevity, we omit the domain $\mathbb{R}^2$ in the following integrals.

During transit, the sky-plane intensity is given by the sum of planetary emission and stellar light transmitted through the planet:
\begin{align}
I(\bm{x},t)
&= I_{\mathrm{p}}(\bm{x}) + \mathcal{T}_{\mathrm{p}}(\bm{x})\,I_{\mathrm{s}}(\bm{x},t) \notag\\
&= I_{\mathrm{p}}(\bm{x}) + \left[1-A_{\mathrm{p}}(\bm{x})\right] I_{\mathrm{s}}(\bm{x},t).
\label{eq:I_def}
\end{align}
The time derivative of the flux is therefore
\begin{align}
\dot{F}(t)
&= \frac{d}{dt}\iint I(\bm{x},t)\,dx\,dy \notag\\
&= -\frac{d}{dt}\iint I_{\mathrm{s}}(\bm{x},t)\,A_{\mathrm{p}}(\bm{x})\,dx\,dy \notag\\
&= -\iint \frac{\partial}{\partial t} I_{\mathrm{s}}(\bm{x},t)\,A_{\mathrm{p}}(\bm{x})\,dx\,dy,
\label{eq:F_tr_dot}
\end{align}
where we used that the total planetary flux $\iint I_{\mathrm{p}}(\bm{x})\,dx\,dy$ and the total stellar flux $\iint I_{\mathrm{s}}(\bm{x},t)\,dx\,dy$ are constant in time by assumption.
We also assume that differentiation under the integral sign is valid.

We define
\begin{equation}
\nabla_{\bm{X}} \equiv
\left(
\frac{\partial}{\partial X},
\frac{\partial}{\partial Y}
\right)^{\mathrm T}
\end{equation}
in the star-centered frame.
Using Equation~(\ref{eq:Is_loc}), the chain rule gives
\begin{align}
\frac{\partial}{\partial t} I_{\mathrm{s}}(\bm{x},t)
&=
\frac{d}{dt}
I_{\mathrm{s}}^{\star}\!\left(\bm{x}+\bm{X}_{\mathrm{p}}(t)\right) \notag\\
&=
\bm{v}(t)\cdot\nabla_{\bm{X}}\,
I_{\mathrm{s}}^{\star}\!\left(\bm{x}+\bm{X}_{\mathrm{p}}(t)\right),
\label{eq:Is_del_t}
\end{align}
where
\begin{equation}
\bm{v}(t)\equiv \dot{\bm{X}}_{\mathrm{p}}(t)
=
\left(\dot{X}_{\mathrm{p}}(t),\dot{Y}_{\mathrm{p}}(t)\right)^{\mathrm T}.
\end{equation}

\subsection{Radon-Transform Limit for Transits}
\label{sec:radon_limit_transit}

We now adopt a simplified limit in which the stellar intensity can be locally approximated by a uniform half-plane.
We use the subscripts $\mathrm{i}$ and $\mathrm{e}$ to denote ingress and egress, respectively.
For $j\in\{\mathrm{i},\mathrm{e}\}$, let $t_j$ be the time at which the planet center crosses the stellar limb, so that $\bm{X}_{\mathrm p}(t_j)$ lies on the stellar limb.
Let $\theta_j$ be the angle of the inward normal to the tangent line of the stellar limb at $\bm{X}_{\mathrm p}(t_j)$, measured counterclockwise from the positive $X$-axis.
Let
\begin{equation}
\bm{n}_j \equiv (\cos\theta_j,\sin\theta_j)^{\mathrm T}
\end{equation}
be the corresponding inward unit normal.

\begin{figure}[ht!]
\epsscale{1.1}
\plotone{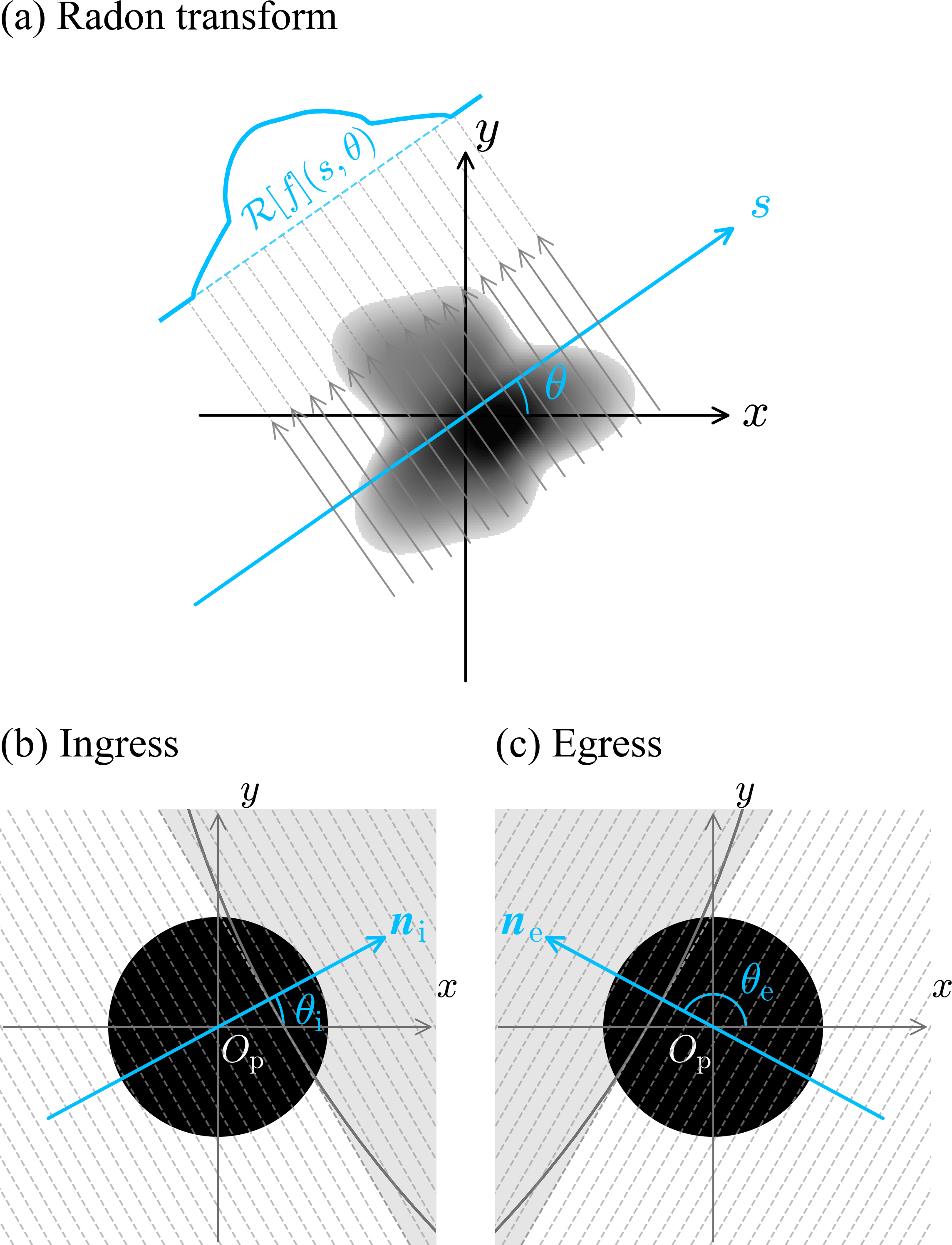}
\caption{(a) Schematic illustration of the Radon transform.
(b,c) Close-up views near the stellar limb during (b) ingress and (c) egress.
In the knife-edge approximation, the curved stellar limb is locally approximated by a straight edge; the dashed lines indicate the locally linearized limb at different times.
The blue arrows indicate the local inward normals to the stellar limb.
In the Radon-transform limit, these normals define the projection-normal direction.
\label{fig:radon_schematic}}
\end{figure}

The uniform half-plane approximation consists of two assumptions:
(i) the stellar limb is locally approximated by a straight knife edge near ingress and egress (Figure~\ref{fig:radon_schematic}b,c), and
(ii) the stellar surface brightness is uniform, so limb darkening is neglected.
The knife-edge approximation is accurate when $R_{\mathrm{p}} \ll R_{\mathrm{curv},j}$, where $R_{\mathrm{p}}$ is a characteristic size scale of the planetary silhouette and $R_{\mathrm{curv},j}$ is the local radius of curvature of the stellar limb near ingress or egress.
The leading curvature-induced normalized flux error scales as
$\sim (R_{\mathrm{p}}/R_{\mathrm{s}})^2
(R_{\mathrm{p}}/R_{\mathrm{curv},j})$, where $R_{\mathrm{s}}$ is a characteristic size scale of the stellar disk, as quantified in Section~\ref{sec:model_comparison}.
We further restrict attention to cases in which the entire planetary silhouette is fully superimposed on the stellar disk at some point during the transit.

Near ingress or egress, we approximate the stellar intensity by
\begin{equation}
I_{\mathrm{s}}^{\star}(\bm{X}) \simeq I_0\,
H\!\left(\bm{n}_j\cdot\left(\bm{X}-\bm{X}_{\mathrm{p}}(t_j)\right)\right),
\label{eq:Is_knife_local}
\end{equation}
where $I_0$ is the uniform stellar surface brightness and $H$ is the Heaviside step function,
\begin{equation}
H(x)=
\begin{cases}
1, & x>0,\\
0, & x<0.
\end{cases}
\end{equation}
Its derivative is understood in the sense of distributions, so that
\begin{equation}
\nabla H(g)=\delta(g)\nabla g,
\end{equation}
where $\delta$ denotes the Dirac delta distribution.
Equation~(\ref{eq:Is_knife_local}) therefore implies
\begin{equation}
\nabla_{\bm{X}} I_{\mathrm{s}}^{\star}(\bm{X}) \simeq 
I_0\,\delta\!\left(\bm{n}_j\cdot\left(\bm{X}-\bm{X}_{\mathrm{p}}(t_j)\right)\right)
\bm{n}_j.
\label{eq:grad_Is_ingress}
\end{equation}

Substituting Equation~(\ref{eq:grad_Is_ingress}) into Equations~(\ref{eq:Is_del_t}) and (\ref{eq:F_tr_dot}) gives
\begin{align}
\dot{F}_j(t)
=
&-I_0\,\bm{n}_j\cdot\bm{v}(t)
\iint
A_{\mathrm{p}}(\bm{x}) \notag\\
&\ \times \delta\!\left(
\bm{n}_j\cdot\left(\bm{x}+\bm{X}_{\mathrm{p}}(t)-\bm{X}_{\mathrm{p}}(t_j)\right)
\right)
\,dx\,dy.
\label{eq:F_tr_del_t_deltaf}
\end{align}

We define the Radon transform of a two-dimensional function $f(\bm{x})$ as \citep{Radon_1917,Radon_eng_1917}
\begin{equation}
\mathcal{R}[f](s,\theta)
=
\iint_{\mathbb{R}^2}
f(\bm{x})\,
\delta\!\left(\bm{n}(\theta)\cdot\bm{x}-s\right)
\,dx\,dy,
\label{eq:Radon}
\end{equation}
where
\begin{equation}
\bm{n}(\theta)\equiv (\cos\theta,\sin\theta)^{\mathrm T}.
\end{equation}
A schematic illustration is shown in Figure~\ref{fig:radon_schematic}a.
Defining
\begin{equation}
s_j(t)
\equiv
-\bm{n}_j\cdot\left(\bm{X}_{\mathrm{p}}(t)-\bm{X}_{\mathrm{p}}(t_j)\right),
\label{eq:sj_def}
\end{equation}
and comparing Equations~(\ref{eq:F_tr_del_t_deltaf}) and (\ref{eq:Radon}), we obtain
\begin{equation}
\dot{F}_j(t)
=
-I_0\,\bm{n}_j\cdot\bm{v}(t)\,
\mathcal{R}[A_{\mathrm{p}}]\!\left(s_j(t),\theta_j\right).
\label{eq:Fdot_tr_Radon}
\end{equation}

Equation~(\ref{eq:Fdot_tr_Radon}) shows that, in this approximation, the time derivative of the flux during ingress or egress traces a single Radon-transform projection of the planetary attenuation map $A_{\mathrm{p}}$.
Ingress and egress provide two Radon projections, generally at different angles.
The orbital motion determines how time parametrizes the Radon offset $s_j(t)$ and contributes the prefactor $-I_0\,\bm{n}_j\cdot\bm{v}(t)$.

\subsection{Comparison with Existing Transit Light-Curve Models}
\label{sec:model_comparison}

\begin{figure*}[t!]
\epsscale{1.14}
\plotone{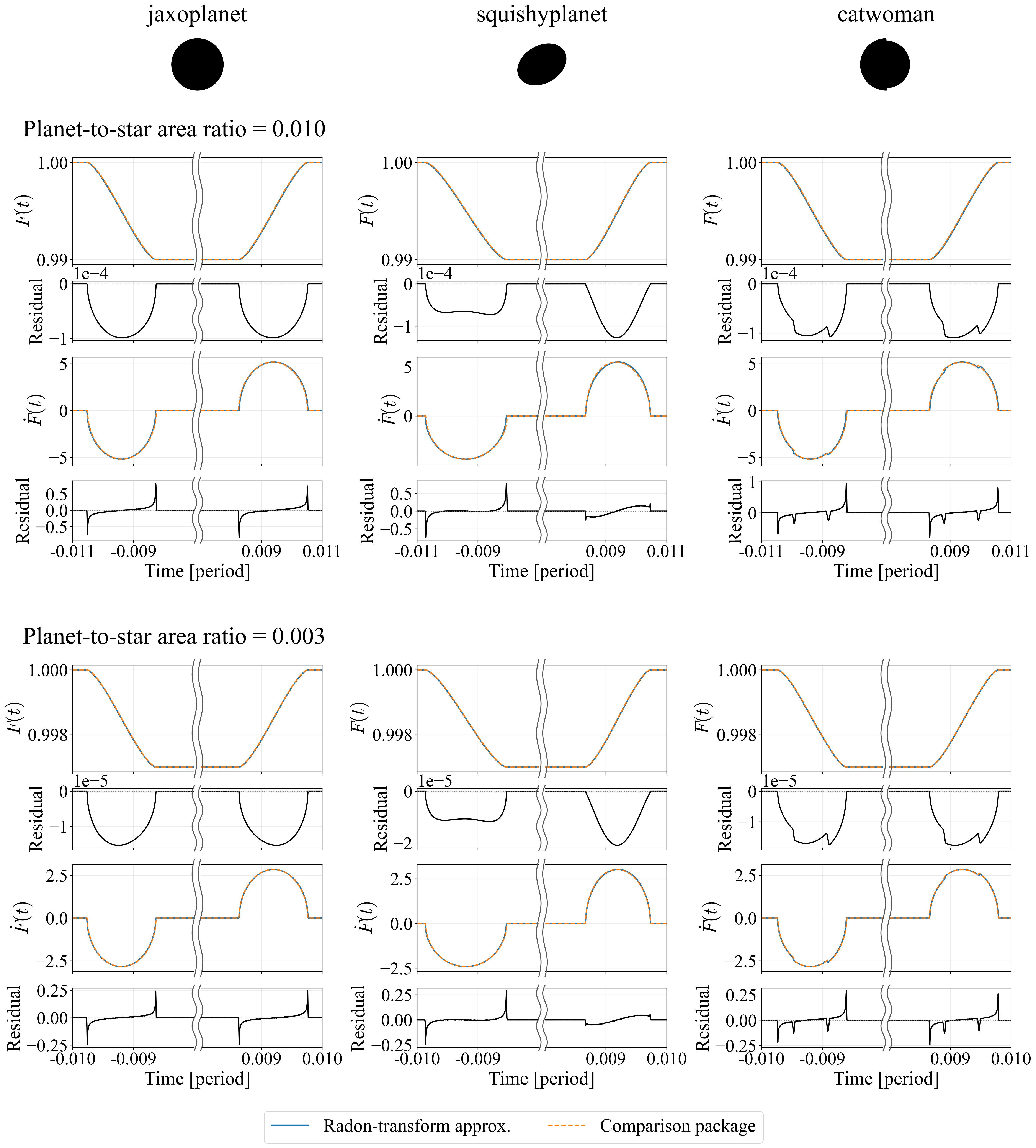}
\caption{
Comparison of light curves from several transit model codes with those predicted by the Radon-transform approximation. Only the portions of the light curves around ingress and egress are shown.
We show cases with planet-to-star area ratios of 0.01 (upper panels) and 0.003 (lower panels).
From left to right, the columns show results for a circular silhouette computed with \texttt{jaxoplanet} \citep{jaxoplanet}, an elliptical silhouette computed with \texttt{squishyplanet} \citep{squishyplanet}, and a silhouette composed of two semicircles computed with \texttt{catwoman} \citep{2021AJ....162..165E,Jones2022}, together with the corresponding Radon-transform approximations.
The planetary silhouettes used in each model are shown at the top of each column.
For each planet-to-star area ratio, the upper two rows show the light curves and their residuals, defined as approximation minus model, while the lower two rows show the corresponding time derivatives and their residuals.
In these calculations, the orbital eccentricity was set to zero, the scaled semi-major axis was set to $a/R_{\mathrm{s}}=15$, the impact parameter was set to 0.5, and the mid-transit time was set to 0.
All model light curves were computed for a uniform stellar disk, without limb darkening.
\label{fig:radon_vs_models}}
\end{figure*}

Figure~\ref{fig:radon_vs_models} shows that the Radon-transform approximation reproduces the model light curves well for the representative examples shown here.
In these calculations, the two adopted planet-to-star area ratios were 0.01 and 0.003.
These area ratios correspond to $R_{\mathrm{p}}/R_{\mathrm{s}}\sim0.10$ and $\sim0.055$, respectively. 
The former is close to a Jupiter-sized planet transiting a Sun-like star,
while the latter corresponds, for example, to an Earth-sized planet transiting a late M dwarf.
For comparison, an Earth--Sun system has a much smaller area ratio, $\sim 8.4\times10^{-5}$.

For the examples shown here, the light-curve residuals remain within about 100~ppm for a planet-to-star area ratio of 0.01 and within about 20~ppm for a ratio of 0.003.
The remaining discrepancy is caused by neglecting stellar-limb curvature.
For a convex stellar disk, the tangent half-plane used in the knife-edge approximation contains a small region outside the true curved stellar disk.
The approximation therefore slightly overestimates the occulted stellar flux during ingress and egress, producing a lower approximate flux and hence predominantly negative residuals when the residuals are defined as approximation minus model.
Equivalently, the approximation places ingress slightly too early and egress slightly too late.

The curvature correction derived in Section~\ref{sec:edge_curvature} and Appendix~\ref{app:curvature_corrections} quantifies this effect through a curvature-corrected offset and additional tangential-moment terms.
We do not apply these higher-order corrections here, because our aim is to clarify the information content of the light curves rather than to construct a higher-accuracy light-curve approximation.

The magnitude of the residuals is consistent with the expected scaling of the curvature error.
For a convex stellar limb, the tangent half-plane used in the knife-edge approximation contains a small region outside the true stellar disk.
Let $\eta$ denote the coordinate along the tangent to the stellar limb, measured from the point of tangency. Assuming that the tangential extent of the silhouette satisfies $|\eta|\leq R_{\mathrm{p}}$, the extra area scales as
\begin{equation}
\Delta A_{\mathrm{curv}}
\lesssim
\int_{-R_{\mathrm{p}}}^{R_{\mathrm{p}}}
\frac{\eta^2}{2R_{\mathrm{curv},j}}\,d\eta
=
\frac{R_{\mathrm{p}}^3}{3R_{\mathrm{curv},j}} .
\end{equation}
Thus the corresponding fractional flux error is
\begin{equation}
\left|\frac{\Delta F}{F_{\mathrm{out}}}\right|_{\mathrm{curv}}
\lesssim
\frac{1}{3\pi}\left(\frac{R_{\mathrm{p}}}{R_{\mathrm{s}}}\right)^2
\left(\frac{R_{\mathrm{p}}}{R_{\mathrm{curv},j}}\right),
\end{equation}
where $F_{\mathrm{out}}$ is the out-of-transit flux.
For a circular stellar disk, this becomes
\begin{equation}
\left|\frac{\Delta F}{F_{\mathrm{out}}}\right|_{\mathrm{curv}}
\lesssim
106~{\rm ppm}
\left[
\frac{R_{\mathrm{p}}/R_{\mathrm{s}}}{0.1}
\right]^{3}.
\end{equation}
This estimate gives about $106$ ppm for $R_{\mathrm{p}}/R_{\mathrm{s}}=0.1$ and about $17$ ppm for $R_{\mathrm{p}}/R_{\mathrm{s}}=0.055$, consistent with the residuals in Figure~\ref{fig:radon_vs_models}.

Overall, the agreement supports the Radon-transform picture as a useful leading-order description of transit light curves.

\section{Information Content and Null Space in the Fourier Plane}
\label{sec:nullspace}

\begin{figure*}[ht!]
\epsscale{1.1}
\plotone{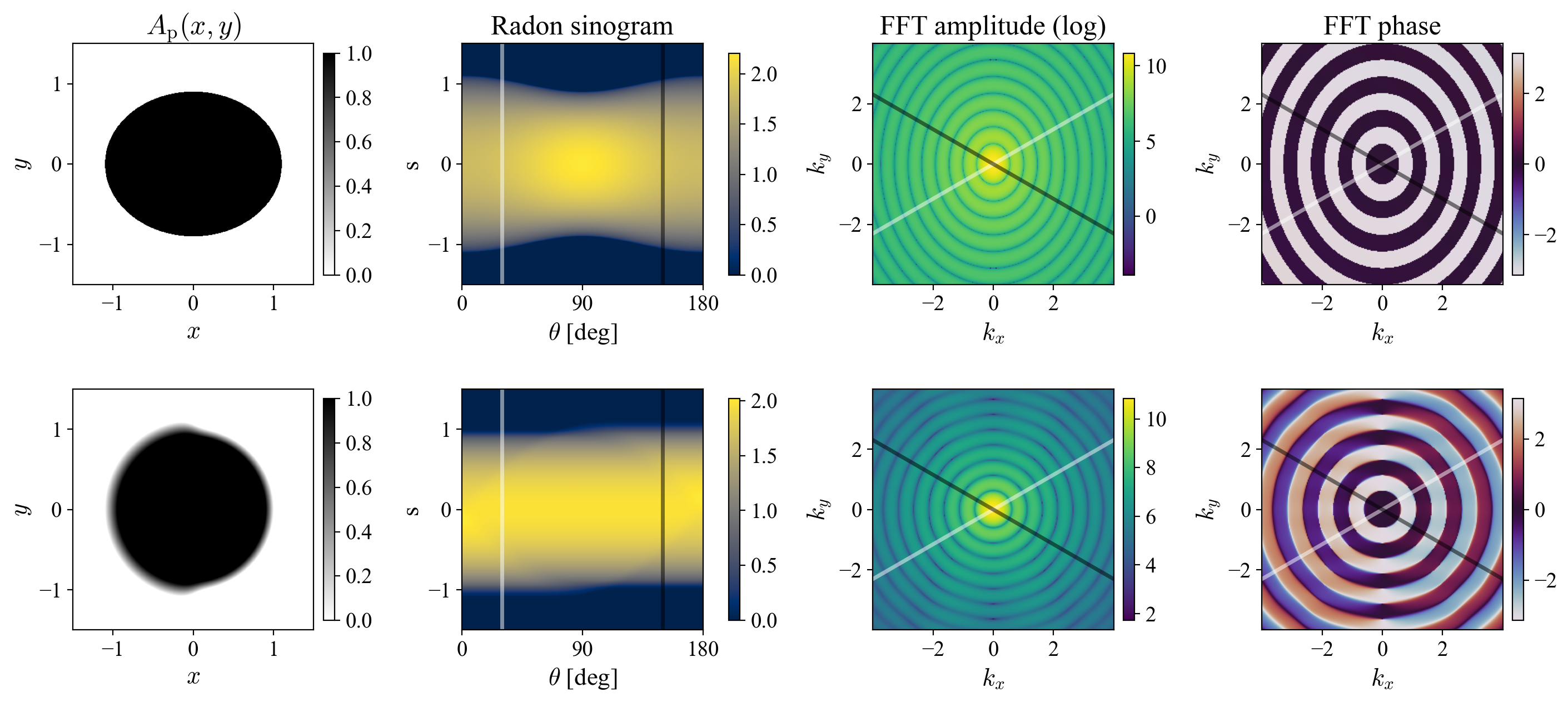}
\caption{Radon projections and Fourier-domain representations of two synthetic spatial patterns of $A_{\mathrm{p}}(x,y)$.
Each row shows, from left to right, the spatial map, its Radon sinogram, the log-scaled Fourier amplitude, and the Fourier phase.
The white and black lines indicate the information accessible from a transit with $\theta_{\mathrm{i}}=30^\circ$ and $\theta_{\mathrm{e}}=150^\circ$, corresponding to an impact parameter of $b \sim 0.5$.
All other regions in the Radon domain and in the Fourier domain are inaccessible in the Radon-transform limit.
\label{fig:radon_fft}}
\end{figure*}

We now turn to the inverse problem of inferring the planetary attenuation map $A_{\mathrm{p}}(\bm{x})$ from transit observations.
As shown above, in the knife-edge, uniform-brightness limit, the time derivative of the light curve provides at most two Radon-transform projections of the underlying two-dimensional map.
Consequently, the available information is fundamentally incomplete, and the inverse problem is ill-posed.
A closely analogous statement holds for secondary eclipses; see Appendix~\ref{app:secondary_eclipse}.

\subsection{Fourier-Slice Theorem and Accessible Modes}
\label{sec:fourier_slices}

In the Radon-transform limit, the structure of this incompleteness becomes particularly transparent in the Fourier domain.
The key mathematical tool is the Fourier-slice theorem \citep{1956AuJPh...9..198B}.
For a two-dimensional function $f(\bm{x})$, we define its Fourier transform as
\begin{equation}
\widehat{f}(k_x,k_y)
\equiv
\iint_{\mathbb{R}^2}
f(x,y)\,
e^{-2\pi i(k_x x+k_y y)}\,dx\,dy,
\label{eq:FT2_def}
\end{equation}
while for a one-dimensional function $g(s)$ we define
\begin{equation}
\widetilde{g}(k)
\equiv
\int_{\mathbb{R}} g(s)\,e^{-2\pi i k s}\,ds.
\label{eq:FT1_def}
\end{equation}

The Fourier-slice theorem states that the one-dimensional Fourier transform of the Radon transform $\mathcal{R}[f](s,\theta)$ with respect to the offset $s$ is equal to a slice of the two-dimensional Fourier transform of $f$:
\begin{align}
\widetilde{\mathcal{R}[f]}(k,\theta)
&\equiv
\int_{\mathbb{R}}
\mathcal{R}[f](s,\theta)\,
e^{-2\pi i k s}\,ds \notag\\
&=
\widehat{f}\!\left(k\cos\theta,\,k\sin\theta\right).
\label{eq:fourier_slice}
\end{align}
Therefore, for a fixed projection angle $\theta$, Radon-transform measurements constrain $\widehat{f}(k_x,k_y)$ only along a single radial line in the Fourier plane.

In the context of a transit, each of ingress and egress provides one such projection angle.
As a result, the observational data constrain $\widehat{A_{\mathrm{p}}}$ on at most two radial lines.
All remaining values in the Fourier plane cannot be recovered without introducing additional assumptions or information.
This structure is illustrated in Figure~\ref{fig:radon_fft}, which shows the accessible Fourier slices and the corresponding inaccessible regions.

To connect temporal Fourier frequencies to spatial Fourier modes, it is convenient to make one additional approximation: over each of ingress and egress, the projected velocity is approximately constant.
We therefore write
\begin{equation}
\bm{X}_{\mathrm{p}}(t)
\simeq
\bm{X}_{\mathrm{p}}(t_j)+\bm{v}(t_j)(t-t_j),
\label{eq:const_velocity}
\end{equation}
and define
\begin{equation}
\alpha_j \equiv \bm{n}_j\cdot\bm{v}(t_j).
\end{equation}
Equation~(\ref{eq:sj_def}) then becomes
\begin{equation}
s_j(t)\simeq -\alpha_j(t-t_j),
\end{equation}
so Equation~(\ref{eq:Fdot_tr_Radon}) reduces to
\begin{equation}
\dot{F}_j(t)
\simeq
-I_0\,\alpha_j\,
\mathcal{R}[A_{\mathrm{p}}]\!\left(-\alpha_j(t-t_j),\theta_j\right).
\label{eq:Fdot_linearized_time}
\end{equation}

Taking the Fourier transform of Equation~(\ref{eq:Fdot_linearized_time}) with respect to time gives
\begin{align}
\widetilde{\dot{F}_j}(\nu)
&=
\int_{\mathbb{R}} \dot{F}_j(t)\,e^{-2\pi i\nu t}\,dt \notag\\
&=
-\sigma_j I_0
e^{-2\pi i\nu t_j}
\widetilde{\mathcal{R}[A_{\mathrm{p}}]}\!\left(-\frac{\nu}{\alpha_j},\theta_j\right) \notag\\
&=
-\sigma_j I_0
e^{-2\pi i\nu t_j}
\widehat{A_{\mathrm{p}}}\!\left(
-\frac{\nu\cos\theta_j}{\alpha_j},
-\frac{\nu\sin\theta_j}{\alpha_j}
\right),
\label{eq:tr_Radon_FT}
\end{align}
where $\nu$ is the temporal frequency and
\begin{equation}
\sigma_j \equiv \frac{\alpha_j}{|\alpha_j|}.
\label{eq:fourier_sigma}
\end{equation}
This expression shows explicitly that $\widetilde{\dot{F}_{\mathrm{i}}}$ and $\widetilde{\dot{F}_{\mathrm{e}}}$ sample $\widehat{A_{\mathrm{p}}}$ only along two radial lines in the Fourier plane.

\subsection{Null Space and Non-Uniqueness}
\label{sec:nullspace_nonuniqueness}

In this idealized limit, the null space of the forward operator from $A_{\mathrm{p}}(\bm{x})$ to the ingress/egress signal consists of all functions whose Fourier transforms vanish on the observed radial slices.
That is, the null space includes all maps $f(\bm{x})$ satisfying
\begin{equation}
\widehat{f}\!\left(k\cos\theta,\,k\sin\theta\right)=0,
\qquad
\forall k\in\mathbb{R},\ 
\theta\in\{\theta_{\mathrm{i}},\,\theta_{\mathrm{e}}\}.
\label{eq:nullspace_fourier_condition_Ap}
\end{equation}
Consequently, the ingress/egress signals generated by the true attenuation map $A_{\mathrm{p}}(\bm{x})$ are indistinguishable from those generated by $A_{\mathrm{p}}(\bm{x})+f(\bm{x})$ whenever $f(\bm{x})$ satisfies Equation~(\ref{eq:nullspace_fourier_condition_Ap}).
In practice, only perturbations that keep the resulting map physically admissible correspond to viable alternative solutions.

Physically motivated priors reduce the feasible set but do not, in general, eliminate the non-uniqueness.
For attenuation maps, it is natural to impose
\begin{equation}
0 \le A_{\mathrm{p}}(\bm{x}) \le 1.
\end{equation}
For a strictly opaque silhouette, one may further impose the stronger binary constraint
\begin{equation}
A_{\mathrm{p}}(\bm{x})\in\{0,1\}.
\end{equation}
One may also assume compact support or a low-dimensional parametric family, such as circles, ellipses, or low-order boundary perturbations.
Such assumptions reduce the range of admissible solutions, but they may not be sufficient to eliminate the non-uniqueness.

Spectroscopic observations provide another way to restrict the set of physically acceptable solutions, although they do not by themselves remove the geometric null space.
In the Radon-transform limit, multi-wavelength data sample $A_{\mathrm p}(\bm{x},\lambda)$, where $\lambda$ denotes wavelength, along the same ingress and egress projection angles at each wavelength. Thus, if the maps are allowed to vary independently at each wavelength, the geometric null space associated with the limited angular coverage remains.
However, physically motivated atmospheric models can constrain the wavelength dependence of $A_{\mathrm p}(\bm{x},\lambda)$.
Solutions that require physically implausible transmission spectra can therefore be excluded.

\begin{figure}[ht!]
\centering
\includegraphics[width=\linewidth]{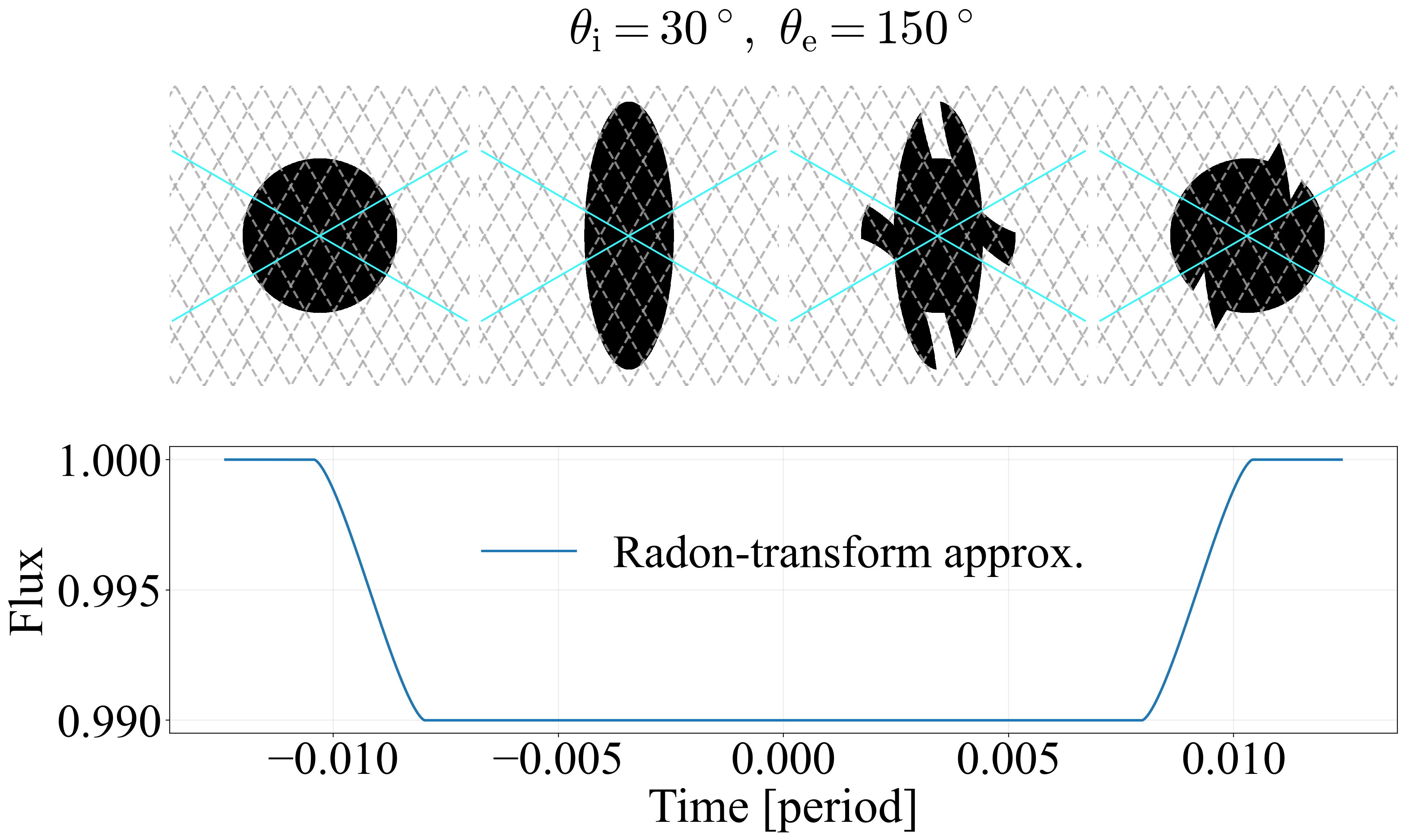}\\
\vspace{3mm}
\includegraphics[width=\linewidth]{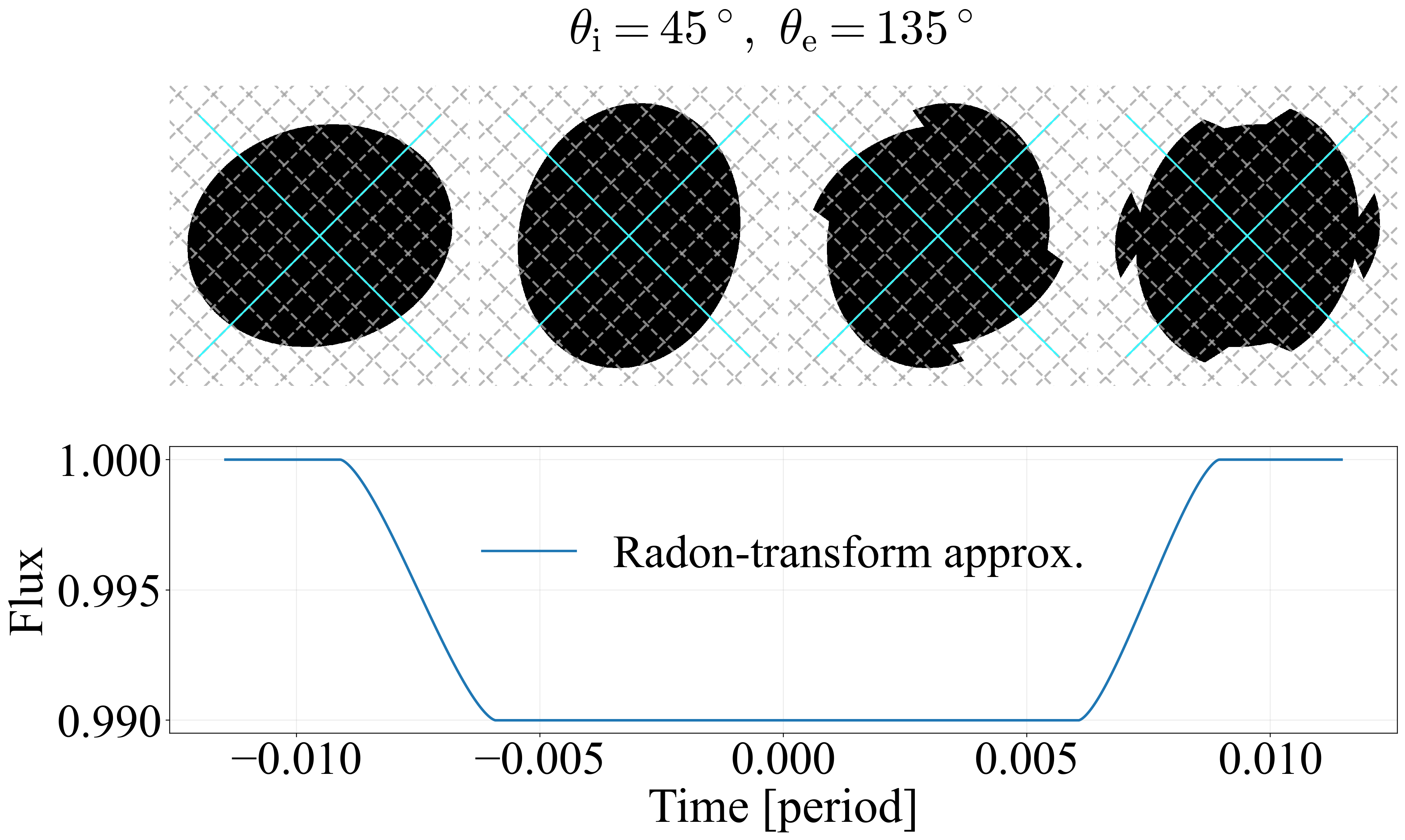}
\caption{
Examples of non-uniqueness for opaque silhouettes in the Radon-transform limit.
The four panels in the top row show silhouettes that produce the same light curve for $(\theta_{\mathrm{i}}, \theta_{\mathrm{e}}) = (30^{\circ}, 150^{\circ})$: a circular silhouette, an elliptical silhouette, and two more complicated shapes.
The four panels in the third row show silhouettes that produce the same light curve for $(\theta_{\mathrm{i}}, \theta_{\mathrm{e}}) = (45^{\circ}, 135^{\circ})$: two elliptical silhouettes with different projected position angles and two more complicated shapes.
The blue lines indicate the local normals to the stellar limb at ingress and egress, and the gray dashed lines show a grid parallel to the locally linearized limb.
The corresponding light curves are shown beneath each group of silhouettes.
In these calculations, the planet-to-star area ratio was set to 0.01, the orbital eccentricity to zero, the scaled semi-major axis to $a/R_{\mathrm{s}}=15$, and the mid-transit time to 0. The impact parameter was calculated to achieve the given $(\theta_{\mathrm{i}}, \theta_{\mathrm{e}})$.
\label{fig:shape_degeneracy}}
\end{figure}

Figure~\ref{fig:shape_degeneracy} shows examples of non-uniqueness in the Radon-transform limit. In each set of four panels, the opaque silhouettes produce identical light curves for the indicated pair of angles.
For $(\theta_{\mathrm{i}}, \theta_{\mathrm{e}}) = (30^{\circ}, 150^{\circ})$, we show silhouettes that produce the same light curve as a circular silhouette, including an elliptical silhouette and two more complicated shapes.
For $(\theta_{\mathrm{i}}, \theta_{\mathrm{e}}) = (45^{\circ}, 135^{\circ})$, we show two elliptical silhouettes with different projected position angles and two more complicated shapes that produce identical light curves.
These examples illustrate that the silhouette cannot be uniquely inferred from the light curve alone.
Accordingly, any inferred silhouette should be understood as the result of combining the data with the adopted prior structure.

\subsection{Degeneracies with Orbital Parameters}
\label{sec:shape_orbit_degeneracy}

The discussion above assumes that the mapping from observation time to Radon offset is known.
In practice, however, that mapping depends on the orbital geometry and, more generally, on the local geometry of the stellar limb sampled by the transit.
For a spherical star, the projection angles are determined mainly by the impact parameter $b$, while the conversion from time to Radon offset depends mainly on the quantity $\bm{n}_j\cdot\bm{v}(t_j)$.
Thus, the spatial information encoded in the transit light curve cannot be interpreted without some knowledge of the orbit.
This introduces an additional degeneracy between the inferred attenuation map and the orbital parameters.
Related degeneracies have been emphasized in earlier studies of oblateness \citep{Barnes_2003,Berardo_2022,Liu_2024,Dholakia_2025}, in more flexible or non-parametric descriptions of transit shadows \citep{Sandford_2019,Arkhypov_2021,Nachmani_2022}, and in eclipse mapping \citep{deWit_2012,2018AJ....156..235R,2024ApJ...969L..32C}.

The interval $t_{\mathrm{e}}-t_{\mathrm{i}}$ provides some information on the orbital parameters, but not enough to remove this degeneracy.
For a circular orbit around a spherical star,
\begin{align}
t_{\mathrm{i}}
&=
t_0-\frac{P}{2\pi}\arcsin\!\left(\sqrt{\frac{1-b^2}{a^2/R_{\mathrm{s}}^2-b^2}}\right), \label{eq:ti_circular} \\
t_{\mathrm{e}}
&=
t_0+\frac{P}{2\pi}\arcsin\!\left(\sqrt{\frac{1-b^2}{a^2/R_{\mathrm{s}}^2-b^2}}\right),
\label{eq:te_circular}
\end{align}
where $t_0$ is the mid-transit time, $P$ is the orbital period, $a$ is the semi-major axis, and $R_{\mathrm{s}}$ is the stellar radius.
Thus, the interval $t_{\mathrm{e}}-t_{\mathrm{i}}$ constrains only a combination of $P$, $a/R_{\mathrm{s}}$, and $b$.
Because $b$ is difficult to constrain independently of the transit light curve, external constraints on the orbital and stellar parameters are needed to infer $b$ from this interval.
For eccentric orbits, the situation is even less constrained, because the eccentricity and the argument of periastron also affect the interval.

Accordingly, without external information, the spatial interpretation of ingress and egress remains strongly degenerate with the orbital solution.
Independent constraints on the orbital parameters are therefore needed to reduce this degeneracy.
If multiple transits are observed, the orbital period can be determined precisely.
If secondary eclipses or radial-velocity measurements are available, they can help constrain the eccentricity and the argument of periastron.
Likewise, the stellar density can be constrained through stellar modeling using available observational constraints, including spectroscopy, photometry, parallax, and asteroseismic information.
Under the usual assumptions of Keplerian motion and neglecting the planetary mass relative to the stellar mass, it is related to $a/R_{\mathrm{s}}$ through
\begin{equation}
\frac{a}{R_{\mathrm{s}}}
=
\left(
\frac{G P^2 \rho_{\mathrm{s}}}{3\pi}
\right)^{1/3},
\label{eq:aRs_rhos}
\end{equation}
where $\rho_{\mathrm{s}}$ is the stellar density and $G$ is the gravitational constant.

More generally, silhouette inference should be understood as depending not only on the light curve itself, but also on prior information about the orbital and stellar properties.

\section{Beyond the Radon-Transform Limit}
\label{sec:deviations}

We now relax the idealized assumptions adopted above and examine how realistic effects modify the Radon-transform limit and its associated null space.
Below, we discuss four important examples: stellar limb darkening, finite exposure time, curvature of the stellar limb, and time-dependent planetary maps.

\subsection{Limb Darkening}
\label{sec:limb_darkening}

When stellar limb darkening is included, the stellar intensity is no longer a uniform half-plane near the limb.
Nevertheless, in the local knife-edge limit, assuming that the stellar intensity varies primarily along the inward-normal direction provides a useful approximation for understanding its effect on the transit signal.
As derived in Appendix~\ref{app:limb_darkening}, the effect of limb darkening is to replace the ideal Radon measurement by a one-dimensional weighted superposition of nearby Radon offsets.

For ingress and egress, the light-curve derivative can therefore be written as
\begin{equation}
\dot{F}_{j}(t) = -\bm{n}_j\cdot \bm{v}(t) \int_{\mathbb R} \Psi_j'(p)\, \mathcal{R}[A_{\mathrm p}] \!\left(s_j(t)+p,\, \theta_j \right)\,dp,
\label{eq:Fdot_limbdark_general_main}
\end{equation}
where $\Psi_j(u)$ is the local stellar brightness profile as a function of distance from the locally linearized limb and $\Psi_j'(u)\equiv d\Psi_j(u)/du$.
Thus, under the local one-dimensional brightness approximation, limb darkening does not introduce new projection angles and only mixes neighboring offsets along the same projection direction.

If we further assume, as in Section~\ref{sec:fourier_slices}, that the projected velocity is approximately constant over ingress or egress, the convolution can be written directly in time.
Defining
\begin{equation}
K_j(q)\equiv |\alpha_j|\,\Psi_j'(\alpha_j q),
\label{eq:Kj_limbdark_main}
\end{equation}
where $\alpha_j=\bm{n}_j\cdot\bm{v}(t_j)$, Equation~(\ref{eq:Fdot_limbdark_general_main}) becomes
\begin{equation}
\dot{F}_{j}(t)\simeq -\alpha_j\int_{\mathbb R}K_j(q)\,\mathcal{R}[A_{\mathrm p}]
\!\left(-\alpha_j(t-t_j-q),\,\theta_j\right)\,dq .
\label{eq:Fdot_limbdark_conv_main}
\end{equation}
Equivalently, the effect of limb darkening can be viewed as a one-dimensional convolution of the ideal Radon-limit signal with a kernel determined by the local limb-darkening profile.

Taking the temporal Fourier transform gives
\begin{align}
\widetilde{\dot{F}_{j}}(\nu)
&=
-\sigma_j e^{-2\pi i\nu t_j}\widetilde{K_j}(\nu)\,\widehat{A_{\mathrm p}}
\!\left(
-\frac{\nu\cos\theta_j}{\alpha_j},
-\frac{\nu\sin\theta_j}{\alpha_j}
\right),
\label{eq:Fdot_limbdark_FT_main}
\end{align}
where $\widetilde{K_j}$ is the one-dimensional Fourier transform of $K_j$ and
$\sigma_j\equiv \alpha_j/|\alpha_j|$.
The accessible Fourier slices are therefore the same as in the uniform-brightness limit, and limb darkening only reweights them.

Therefore, even if the limb-darkening profile were known perfectly, it would not provide additional information about $A_{\mathrm p}(\bm{x})$ beyond that available in the uniform-brightness limit.
If the limb-darkening profile is inferred simultaneously from the transit light curve, it becomes degenerate with the measurable Radon projections of $A_{\mathrm p}$.
An incorrect limb-darkening model can then bias the inferred attenuation map and hence the recovered planetary silhouette or atmospheric asymmetry.
This is a well-known issue and has been widely discussed in studies of transit light-curve analysis \citep[e.g.,][]{Hellard_2020,Arkhypov_2021,Cassese_2026}.

\subsection{Finite Exposure Time}
\label{sec:finite_exposure}

Real observations do not measure the instantaneous flux.
Instead, each exposure returns a time average over a finite integration time $\Delta t_{\mathrm{exp}}$.
If $F(t)$ denotes the ideal continuous light curve, the exposure-averaged signal is
\begin{align}
F_{\mathrm{avg}}(t)
&=
\frac{1}{\Delta t_{\mathrm{exp}}}
\int_{t-\Delta t_{\mathrm{exp}}/2}^{t+\Delta t_{\mathrm{exp}}/2}
F(\tau)\,d\tau
\notag\\
&\equiv
\left(F * B_{\Delta t_{\mathrm{exp}}}\right)(t),
\label{eq:finite_exposure_def}
\end{align}
where $B_{\Delta t_{\mathrm{exp}}}(t)$ is the unit-area boxcar kernel of width $\Delta t_{\mathrm{exp}}$ and $*$ denotes convolution.
Differentiating gives
\begin{equation}
\dot{F}_{\mathrm{avg}}(t)
=
\left(\dot{F} * B_{\Delta t_{\mathrm{exp}}}\right)(t),
\label{eq:Fdot_smoothed_time}
\end{equation}
so finite exposure smooths the ideal light-curve derivative in time.

Under the same local constant-velocity approximation used in Section~\ref{sec:fourier_slices}, time maps linearly to the Radon offset during each ingress or egress segment:
\begin{equation}
s_j(t)
\simeq
-\alpha_j\,(t-t_j),
\qquad
j\in\{\mathrm{i},\mathrm{e}\},
\label{eq:s_of_t_finite}
\end{equation}
where $\alpha_j=\bm{n}_j\cdot\bm{v}(t_j)$.
Therefore, finite exposure is equivalently a smoothing along the Radon-offset direction,
\begin{align}
\mathcal{R}_{\mathrm{avg}}[A_{\mathrm p}](s,\theta_j)
&=
\left(
\mathcal{R}[A_{\mathrm p}](\cdot,\theta_j)
*
B_{\Delta s_j}
\right)(s),
\label{eq:Radon_smoothed_s}
\end{align}
with effective smoothing width
\begin{equation}
\Delta s_j
=
|\alpha_j|\,\Delta t_{\mathrm{exp}}.
\label{eq:delta_s_exp}
\end{equation}

In the Fourier domain, this smoothing becomes multiplication.
With the convention of Equation~(\ref{eq:FT1_def}),
\begin{equation}
\widetilde{B}_{\Delta t_{\mathrm{exp}}}(\nu)
=
\frac{\sin(\pi\nu\Delta t_{\mathrm{exp}})}
{\pi\nu\Delta t_{\mathrm{exp}}},
\label{eq:boxcar_sinc}
\end{equation}
so
\begin{equation}
\widetilde{\dot{F}_{\mathrm{avg}}}(\nu)
=
\widetilde{\dot{F}}(\nu)\,
\frac{\sin(\pi\nu\Delta t_{\mathrm{exp}})}
{\pi\nu\Delta t_{\mathrm{exp}}}.
\label{eq:Fdot_smoothed_freq}
\end{equation}
Through Equation~(\ref{eq:tr_Radon_FT}), this means that finite exposure suppresses high-spatial-frequency components along the accessible Fourier slices.
It therefore reduces sensitivity to fine structure.

A second limitation is that the smoothed signal is sampled only at discrete times.
If the sampling interval between successive exposure midpoints is $\Delta t_{\mathrm{samp}}$, then the data constrain the smoothed light curve only at those discrete times.
In Fourier space, this introduces the usual Nyquist limit.
Temporal frequencies above roughly
\begin{equation}
|\nu|
\gtrsim
\frac{1}{2\Delta t_{\mathrm{samp}}}
\label{eq:nyquist_time}
\end{equation}
cannot be recovered reliably and may alias into lower frequencies.

In the Radon-transform picture, this sampling interval corresponds in real space to an offset spacing
\begin{equation}
\delta s_j
=
|\alpha_j|\,\Delta t_{\mathrm{samp}}
\label{eq:delta_s_samp}
\end{equation}
along the measured Radon projection.
Equivalently, from Equation~(\ref{eq:tr_Radon_FT}), the temporal frequency $\nu$ maps to the spatial frequency $k$ along the accessible Fourier slice through
\begin{equation}
|k|=\frac{|\nu|}{|\alpha_j|}.
\label{eq:knu_mapping}
\end{equation}
Therefore, discrete sampling implies a Nyquist-like upper limit
\begin{equation}
|k|
\lesssim
\frac{1}{2|\alpha_j|\Delta t_{\mathrm{samp}}}
=
\frac{1}{2\delta s_j}
\label{eq:nyquist_spatial}
\end{equation}
on the recoverable spatial frequency along the measured slice.
Spatial frequencies above this scale cannot be recovered reliably and may alias into lower frequencies.
Equivalently, only structures broader than roughly $2\delta s_j$ can be robustly resolved along the measured projection.

For a circular orbit, the sky-projected orbital speed during transit is approximately
$v_\perp\simeq 2\pi a/P$.
For a non-grazing transit with impact parameter $b$, the component of the velocity normal to the local stellar limb is
$|\alpha_j|=|\bm{n}_j\cdot\bm{v}|\simeq v_\perp\sqrt{1-b^2}$.
The finite-exposure smoothing width is therefore
\begin{equation}
\begin{aligned}
\frac{\Delta s_{j}}{R_{\mathrm p}}
&\simeq
2\pi \sqrt{1-b^2}
\frac{a\Delta t_{\mathrm{exp}}}{R_{\mathrm p}P} \\
&\simeq
0.024 \sqrt{1-b^2}
\left(\frac{a}{10R_{\mathrm{s}}}\right)
\left(\frac{R_{\mathrm{p}}}{0.1R_{\mathrm{s}}}\right)^{-1} \\
&\qquad\times
\left(\frac{\Delta t_{\mathrm{exp}}}{10\,\mathrm{s}}\right)
\left(\frac{P}{3\,\mathrm{d}}\right)^{-1}.
\end{aligned}
\end{equation}
Similarly, the sampling-limited spatial scale is
\begin{equation}
\begin{aligned}
\frac{\ell_{\mathrm{samp}}}{R_{\mathrm{p}}}
&\sim
\frac{2\delta s_j}{R_{\mathrm{p}}}
\simeq
4\pi \sqrt{1-b^2}
\frac{a\Delta t_{\mathrm{samp}}}{R_{\mathrm{p}}P} \\
&\simeq
0.05 \sqrt{1-b^2}
\left(\frac{a}{10R_{\mathrm{s}}}\right)
\left(\frac{R_{\mathrm{p}}}{0.1R_{\mathrm{s}}}\right)^{-1} \\
&\qquad\times
\left(\frac{\Delta t_{\mathrm{samp}}}{10\,\mathrm{s}}\right)
\left(\frac{P}{3\,\mathrm{d}}\right)^{-1}.
\end{aligned}
\end{equation}
Thus, for typical hot-Jupiter parameters and JWST-like high-cadence observations with sampling intervals from several seconds to several tens of seconds, the native sampling corresponds to spatial scales of order a few percent to a few tenths of a planetary radius.
In practice, the recoverable scale will usually be limited more strongly by photon noise and instrumental systematics than by the native sampling alone.

\subsection{Stellar-Limb Curvature}
\label{sec:edge_curvature}

As shown in Appendix~\ref{app:curvature_corrections}, once the stellar limb is no longer treated as locally straight, the light-curve derivative is no longer described by a pure Radon projection alone.
In Equation~(\ref{eq:Is_curved_eff_app}), the effect of curvature can be separated into a shift of the effective offset and tangentially dependent corrections around the ideal slice.
Correspondingly, Equation~(\ref{eq:Fdot_curved_moment_app}) shows that $\dot{F}_j(t)$ is no longer determined only by the usual Radon-transform term $M_0[A_{\mathrm p}] = \mathcal{R}[A_{\mathrm p}]$, but also contains terms involving $M_1[A_{\mathrm p}]$, $M_2[A_{\mathrm p}]$, and their derivative with respect to $s$, where $M_m$ denotes the $m$th tangential moment defined in Equation (\ref{eq:Mm_def_app}).
Thus, stellar-limb curvature changes the light-curve derivative from a pure Radon projection into a quantity weakly mixed with information from the neighborhood of that projection.

The Fourier-space interpretation is given by Equation~(\ref{eq:Mm_fourier_app}).
There, $M_0$ corresponds to the value of $\widehat{A}_{\mathrm p}$ on the ideal radial slice, while $M_1$ and $M_2$ correspond to the first and second transverse derivatives about that slice.
Therefore, small stellar-limb curvature does not open genuinely new projection angles.
Instead, it makes the data weakly sensitive to transverse structure in $\widehat{A}_{\mathrm p}$ around the ideal ingress and egress slices, a sensitivity that is absent in the strict Radon-transform limit.

The overlap between the information carried by the ingress and egress measurements is minimized when the two ideal slices are orthogonal, which occurs at $b \sim 1/\sqrt{2}$.
In that case, the region of Fourier space that becomes weakly accessible through curvature effects is broadened in the most complementary way.

\subsection{Time-Dependent Planetary Maps}
\label{sec:time_dependent_maps}
The derivation above assumes that the sky-projected planetary maps are effectively static in the planet-centered frame during a single event.
For transits, the attenuation map $A_{\mathrm p}$ is assumed to be static, and the disk-integrated contribution of $I_{\mathrm p}$ must be constant over the event.
For the secondary-eclipse case summarized in Appendix~\ref{app:secondary_eclipse}, the planetary brightness map $I_{\mathrm p}$ is assumed to be static.
These assumptions may be inaccurate for short-period planets that rotate appreciably during a transit or secondary eclipse.
For example, WASP-121b, WASP-18b, and WASP-43b have orbital periods of order a day and are expected to be tidally locked, implying that a transit or eclipse lasting a few hours corresponds to a rotational-phase change of order tens of degrees
\citep[e.g.,][]{2016MNRAS.458.4025D,2009Natur.460.1098H,2011A&A...535L...7H}.
Such systems provide useful opportunities for probing longitudinal atmospheric structure
\citep{2023Natur.620..292C, 2025NatAs...9.1821C,2024AJ....168....4H,2024ApJ...969L..32C,2026NatAs.tmp..121G}.

If the planetary map is time dependent, the flux derivative contains terms beyond the motion of the stellar limb across a fixed map.
We now write the planetary emission and attenuation maps as
$I_{\mathrm p}(\bm{x},t)$ and $A_{\mathrm p}(\bm{x},t)$.
For a transit, using Equations~(\ref{eq:F_def}) and~(\ref{eq:I_def}), we obtain
\begin{align}
\dot{F}(t)
&=
\iint \frac{\partial I_{\mathrm p}}{\partial t}(\bm{x},t)\,dx\,dy
-\iint I_{\mathrm s}(\bm{x},t)
\frac{\partial A_{\mathrm p}}{\partial t}(\bm{x},t)\,dx\,dy \notag\\
&\quad
-\iint A_{\mathrm p}(\bm{x},t)
\frac{\partial I_{\mathrm s}}{\partial t}(\bm{x},t)\,dx\,dy .
\end{align}
The last term is the limb-motion term and reduces to the Radon-transform expression in the static-map limit, under the same local-limb approximation used above.
The first two terms represent changes in the planetary emission and attenuation maps in the sky-projected frame, including rotation-induced changes or intrinsic atmospheric evolution.

If the map-evolution timescale is long compared with an individual ingress or egress, a useful approximation is to treat the map as static within each event but allow it to differ between ingress and egress.
One may then define
$A_{\mathrm p,i}(\bm{x})\equiv A_{\mathrm p}(\bm{x},t_{\mathrm i})$ and
$A_{\mathrm p,e}(\bm{x})\equiv A_{\mathrm p}(\bm{x},t_{\mathrm e})$.
The Radon-transform expression then applies separately to each event,
\begin{equation}
\dot{F}_j(t)
\simeq
-I_0\,\bm{n}_j\cdot\bm{v}(t)\,
\mathcal{R}[A_{\mathrm p,j}]\!\left(s_j(t),\theta_j\right),
\end{equation}
where $j\in\{\mathrm i,\mathrm e\}$.
In this case, ingress and egress provide one projection of each of two different maps, rather than two projections of the same map.
The inverse problem is therefore even more underconstrained than in the static-map case.
The same considerations apply to secondary eclipses.

The two maps can be related by introducing additional physical assumptions, for example by assuming a common three-dimensional structure fixed in the rotating planetary frame.
Such a structure may include the shape of the planet, such as an oblate or tidally distorted figure, as well as an atmospheric structure constrained by a dynamical model such as a general circulation model.
Such assumptions reduce the effective number of degrees of freedom, but the resulting constraints depend on the adopted dynamical or geometric prior.

\section{Implications for Transit Shape Inference}
\label{sec:shape_inference}

In this section, we discuss several implications of the Radon-transform limit for practical transit light-curve analyses with non-circular planetary silhouettes.

\subsection{Oblateness Inference}
\label{sec:oblateness}

In studies of exoplanet oblateness, it is common to approximate the projected planetary shape as an ellipse \citep[e.g.,][]{Seager_2002,Dholakia_2025,Cassese_2026}.
The Radon-transform limit clarifies what is and is not constrained in this case.

For the examples in this subsection, we choose the $x$-axis parallel to the sky-projected orbital velocity at mid-transit and the $y$-axis perpendicular to it.
Let the projected boundary be an ellipse with semi-axes $(R_{\mathrm{major}}, R_{\mathrm{minor}})$ and position angle $\phi$ on the sky.
We define the projected oblateness as $f=1-R_{\mathrm{minor}}/R_{\mathrm{major}}\geq0$.
The distance from the ellipse center to the tangent line whose inward normal makes angle $\theta$ is
\begin{equation}
h(\theta)
=
\sqrt{
R_{\mathrm{major}}^2 \cos^2(\theta-\phi)
+
R_{\mathrm{minor}}^2 \sin^2(\theta-\phi)
}.
\label{eq:ellipse_support}
\end{equation}
For a centered uniform ellipse, each fixed-angle Radon slice has a universal form and is determined only by the projected area $\pi R_{\mathrm{major}} R_{\mathrm{minor}}$ and by the support value $h(\theta)$.
Thus, if the projected area and the orbital scaling factors relating time to Radon offset are known, ingress and egress constrain the ellipse through $h(\theta_{\mathrm i})$ and $h(\theta_{\mathrm e})$ together with $\pi R_{\mathrm{major}} R_{\mathrm{minor}}$.
In that idealized case, the allowed ellipse parameters are reduced to a finite set of solutions, often with two branches.

\begin{figure}[tb!]
\epsscale{1.15}
\plotone{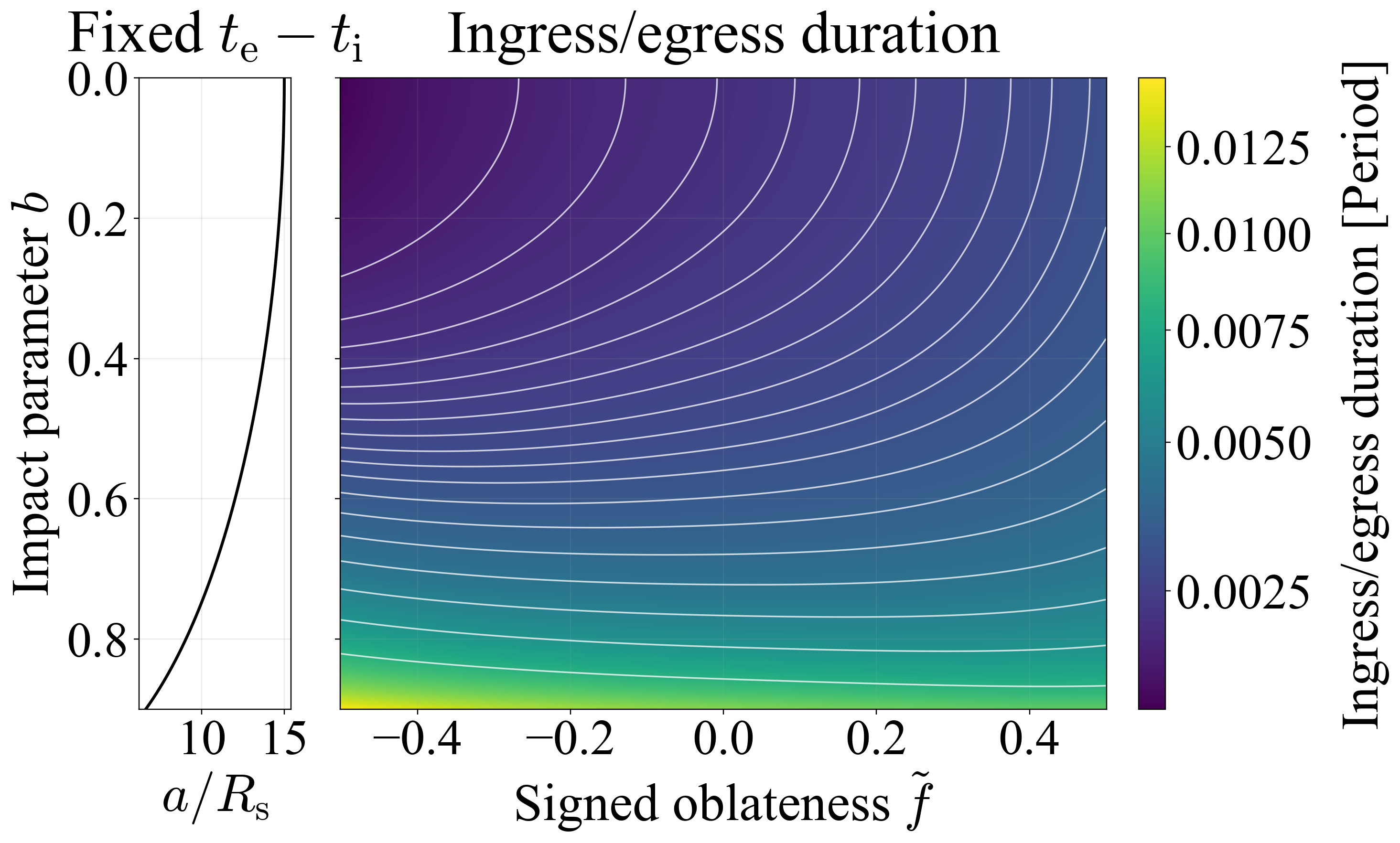}
\caption{Ingress (or egress) duration as a function of the signed oblateness parameter $\tilde{f}$ and the impact parameter $b$ for circular orbits in the Radon-transform limit.
Over each of ingress and egress, the projected velocity is assumed to be constant (Equation~(\ref{eq:const_velocity})).
The projected planetary shape is an ellipse with position angle $\phi=0^\circ$ or $90^\circ$, measured from the $x$-axis.
We define $f=1-R_{\mathrm{minor}}/R_{\mathrm{major}}\geq0$ and use $\tilde{f}=f$ for $\phi=0^\circ$ and $\tilde{f}=-f$ for $\phi=90^\circ$.
This convention allows the projected radii along the $x$- and $y$-axes to vary continuously on a single axis.
The scaled semi-major axis $a/R_{\mathrm{s}}$ was adjusted so that the interval $t_{\mathrm{e}}-t_{\mathrm{i}}$ remained fixed; its value is shown in the left panel.
In these calculations, the planet-to-star area ratio was set to 0.01, and $a/R_{\mathrm{s}}=15$ at $b=0$.
White contours indicate parameter combinations that produce identical light curves in this approximation.
The color scale is non-linear for visual clarity.
\label{fig:ellipse}}
\end{figure}

In practice, however, the time-domain light curve constrains $h(\theta_j)/|\alpha_j|$ rather than $h(\theta_j)$ itself, where $\alpha_j$ is the factor converting time to Radon offset during ingress or egress.
If the projection angles and scaling factors are uncertain because $b$, $a/R_{\mathrm{s}}$, or eccentricity are imperfectly known, then families of elliptical models can produce identical ingress and egress signals in this limit.
Figure~\ref{fig:ellipse} illustrates this explicitly for circular orbits and position angles $\phi=0^\circ$ and $90^\circ$.
Changes in projected oblateness can be compensated by changes in $b$ and $a/R_{\mathrm{s}}$, producing extended degeneracies in the inferred shape parameters.

This viewpoint is closely related to the degeneracies emphasized in earlier oblateness studies \citep[e.g.,][]{Liu_2024,Dholakia_2025}.
External information, such as independent constraints on stellar density, eccentricity, and other orbital parameters, can reduce the allowed parameter space.
At the same time, the inferred oblateness may then depend sensitively on those external constraints.
Conversely, such degeneracies can themselves be used as a search space for non-circular or ringed silhouettes rather than treated only as nuisance parameters \citep{Zuluaga_2015,Arkhypov_2021}.

The perturbative corrections discussed in Section~\ref{sec:edge_curvature} can reduce part of the idealized degeneracy.
Because stellar-limb curvature makes the light curve weakly sensitive to a neighborhood of the ideal Fourier slices, it can help distinguish solutions that are exactly degenerate in the strict Radon-transform limit.
This framework also explains why oblateness constraints are often strongest at relatively high impact parameters.
\citet{Barnes_2003} found that the detectability is maximized near $b\sim0.7$, and \citet{Dholakia_2025} similarly found that systems with $0.7\lesssim b\lesssim0.8$ tend to yield the most unique constraints.
This can be understood as a consequence of ingress and egress probing increasingly complementary projection directions as $b$ increases from 0, with the complementarity peaking near $b\sim1/\sqrt{2}$, before decreasing again toward grazing transits.

\begin{figure}[tb!]
\epsscale{1.15}
\plotone{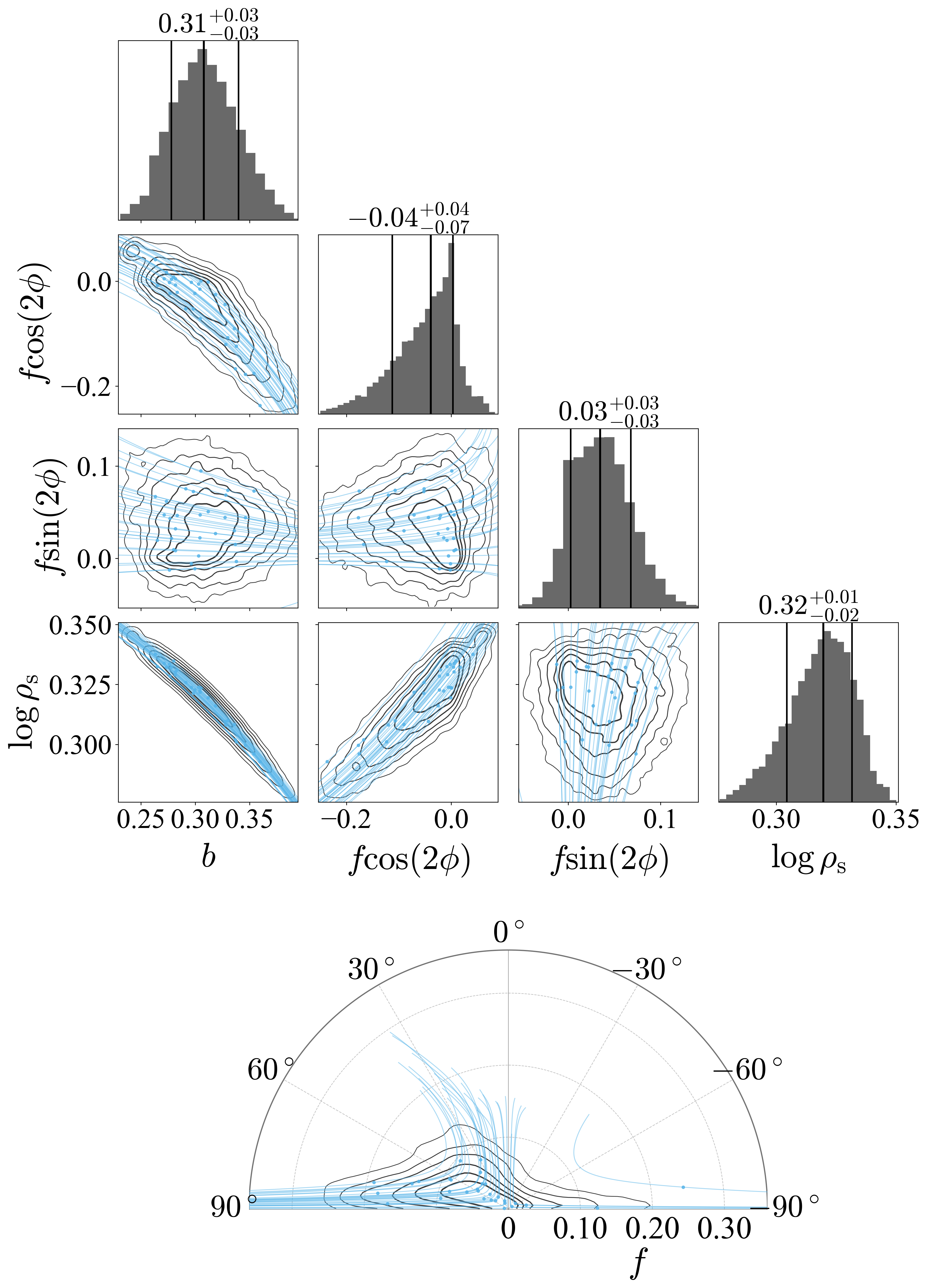}
\caption{
Comparison between the posterior distribution obtained by \citet{Liu_2024} for an elliptical transit model of Kepler-51d and the shape--orbit degeneracy predicted by the Radon-transform approximation.
The upper panels show the posterior in $b$, $f\cos(2\phi)$, $f\sin(2\phi)$, and $\log_{10}[\rho_{\mathrm{s}}/(\mathrm{g\,cm^{-3}})]$.
Here, $f$ and $\phi$ denote the projected oblateness and position
angle, respectively. These correspond to $f_\perp$ and $\theta_\perp$ in \citet{Liu_2024}.
The light-blue points indicate selected posterior samples, and the light-blue curves show the analytic degeneracy curves passing through them.
The lower panel shows the same posterior and analytic curves in polar coordinates, with radius $f$ and angle $\phi$.
\label{fig:Liu2024}}
\end{figure}

Figure~\ref{fig:Liu2024} compares the shape--orbit degeneracy predicted by the Radon-transform approximation with the posterior
obtained by \citet{Liu_2024}. 
They modeled the JWST/NIRSpec transit light curve of Kepler-51d using an elliptical transit model.
Here we use posterior samples from their baseline model. The full
corner plot is shown in their Figure~3.

For each selected reference sample, we calculate the degeneracy curve predicted by the Radon-transform approximation.
At ingress or egress, labeled by $j\in\{\mathrm{i},\mathrm{e}\}$, the duration associated with the local Radon slice is
\begin{equation}
T_j = \frac{2h(\theta_j)}{|\bm{n}_j\cdot\bm{v}(t_j)|},
\end{equation}
where $h(\theta_j)$ is given by Equation~(\ref{eq:ellipse_support}).
To trace the degeneracy curve through a reference sample, we solve
\begin{align}
t_{\mathrm{e}}(b,\rho_{\mathrm{s}}) - t_{\mathrm{i}}(b,\rho_{\mathrm{s}}) &= t_{\mathrm{e},*}-t_{\mathrm{i},*}, \\
T_{\mathrm{i}}(b,\rho_{\mathrm{s}},f,\phi) &= T_{\mathrm{i},*},\\
T_{\mathrm{e}}(b,\rho_{\mathrm{s}},f,\phi) &= T_{\mathrm{e},*},
\end{align}
where starred quantities denote the reference sample, and $t_{\mathrm{i}}$ and $t_{\mathrm{e}}$ are given by Equations~(\ref{eq:ti_circular}) and~(\ref{eq:te_circular}), respectively.
Here $a/R_{\mathrm{s}}$ is calculated from $\rho_{\mathrm{s}}$ using Equation~(\ref{eq:aRs_rhos}), with $P$ fixed to the value adopted by \citet{Liu_2024}.
The area-equivalent radius $R_{\mathrm{mean}}\equiv \sqrt{R_{\mathrm{major}}R_{\mathrm{minor}}}$ and all other parameters are fixed at their values in the reference sample.

Each curve is anchored at one selected sample but is not fitted to the posterior.
Its close alignment with the posterior contours shows that the Radon-transform approximation captures the dominant shape--orbit degeneracy in the oblateness inference.
The finite width and extent of the posterior contours reflect effects beyond the Radon-transform approximation, such as stellar-limb curvature and the adopted priors.

\subsection{Atmospheric Inhomogeneity Inference}
\label{sec:atmospheric_inhomogeneity}

Recent JWST studies have begun to separate the morning and evening transmission spectra of transiting planets, revealing inhomogeneous terminators \citep[e.g.,][]{2024Natur.632.1017E,Murphy_2025}.
In such analyses, spectroscopic light curves are likely to constrain primarily large-scale differences between the two limbs, such as wavelength-dependent shifts of the effective silhouette.
This is because spectroscopic light curves generally have lower signal-to-noise ratios than broadband light curves, and because the effects discussed in Section~\ref{sec:deviations} blur small-scale structure.

From the Radon-transform perspective developed in this paper, ingress and egress constrain projections of asymmetries along two different directions set by the local stellar-limb normals.
This perspective is consistent with approaches that analyze ingress and egress separately to probe such asymmetries, including the method proposed by \citet{Tada_2025}.
For rapidly rotating planets, the sky-projected atmospheric map sampled during ingress need not be the same as that sampled during egress.
Interpreting the two signals as projections of a single static map can therefore bias the inferred atmospheric asymmetry unless the rotation-induced time dependence is modeled.
However, such analyses require careful limb-darkening treatment, because errors in the adopted profile can bias the inferred offset when ingress and egress are modeled separately.
If one focuses only on morning-evening asymmetry, this bias may cancel between ingress and egress in cases where both the true and assumed limb-darkening profiles are symmetric.

As discussed in several studies \citep[e.g.,][]{2024NatAs...8.1562M,Fu_2025}, accurate knowledge of the planetary orbit, such as the mid-transit time $t_0$, is essential for defining the reference point of the atmospheric asymmetry.
In such analyses, the planetary orbit is common across wavelengths, but the centroid of the effective silhouette can differ from the planet's center of mass.
This complicates the separation between orbital geometry and atmospheric asymmetry, and is not yet fully resolved in current analyses.

\section{Conclusion}
\label{sec:conclusion}

In this paper, we showed that, in a simple and physically transparent limit, the time derivative of a transit light curve during ingress or egress can be interpreted as a Radon-transform measurement of the planetary attenuation map.
In this picture, ingress and egress provide at most two projection angles, set by the local normals to the stellar limb.
The corresponding formulation for secondary eclipse is closely analogous, with the attenuation map replaced by the planetary brightness map.

This viewpoint makes the information content of a single transit particularly clear.
In the ideal Radon-transform limit, the data constrain the two-dimensional Fourier transform of the planetary map only along at most two radial slices.
The inverse problem therefore has a large null space.
As a result, the planetary silhouette cannot in general be inferred uniquely from the light curve alone.
Any practical inference must combine the light curve with prior assumptions and with external constraints on the orbital and stellar properties.

We also examined how this picture is modified by several realistic effects.
Under a local one-dimensional approximation, limb darkening acts as a convolution along the Radon-offset direction and does not introduce new projection angles.
Finite exposure time and discrete sampling smooth the signal and limit the recoverable spatial frequencies along the measured slices.
Small stellar-limb curvature does not provide entirely new angular coverage, but it introduces weak sensitivity to transverse Fourier-space structure around the ideal slices that is invisible in the strict Radon-transform limit.
If the planetary map changes during the event, for example because of planetary rotation, applying the Radon-transform interpretation to ingress and egress separately is useful, but additional physical assumptions are needed to relate the corresponding maps.

These results provide a compact framework for understanding what ingress and egress can and cannot reveal about non-circular planetary silhouettes.
This perspective may be useful for interpreting future analyses of planetary oblateness, atmospheric asymmetries, and related eclipse-mapping problems.

\begin{acknowledgments}
We thank Quanyi Liu, Wei Zhu, and Kento Masuda for allowing us to use the posterior samples from their analysis.
We also thank Hajime Kawahara, Shota Miyazaki, and Kohei Miyakawa for helpful discussions and valuable suggestions.
We thank the anonymous referee for insightful comments that greatly improved the manuscript.
This work was supported by JSPS KAKENHI Grant Number 25K23414.
We used ChatGPT (OpenAI) to assist with English-language editing and manuscript organization.
The author reviewed and edited all AI-assisted content and takes full responsibility for the content of this paper.
\end{acknowledgments}

\software{
jaxoplanet (v0.1.0) \citep{jaxoplanet}, 
squishyplanet (v0.3.1) \citep{squishyplanet}, 
catwoman (v1.1.0) \citep{2021AJ....162..165E, Jones2022},
JAX (v0.7.2) \citep{jax2018github},
NumPy (v2.3.3) \citep{harris2020array},
Matplotlib (v3.10.7) \citep{Hunter:2007},
Cartopy (v0.25.0) \citep{Cartopy}
          }

\appendix

\section{Secondary Eclipse: Radon-Transform Limit}
\label{app:secondary_eclipse}

In this Appendix, we summarize the secondary-eclipse counterparts of the transit expressions derived in Section~\ref{sec:formulation}.
For secondary eclipse, the observable quantity is the planetary brightness map rather than the attenuation map, and the corresponding inverse problem forms the basis of eclipse mapping \citep{Williams_2006,Rauscher_2007,Majeau_2012,deWit_2012}.

We use the same coordinates defined in Section~\ref{sec:formulation}.
In the star-centered frame, we define the stellar transmittance map
\begin{equation}
\mathcal{T}_{\mathrm s}^{\star}(\bm{X}),
\end{equation}
with $\mathcal{T}_{\mathrm s}^{\star}=0$ inside the projected stellar disk and $\mathcal{T}_{\mathrm s}^{\star}=1$ outside.
We also define
\begin{equation}
A_{\mathrm s}^{\star}(\bm{X})
\equiv
1-\mathcal{T}_{\mathrm s}^{\star}(\bm{X}).
\end{equation}
Expressed in planet-centered coordinates, these become time dependent through the orbital motion:
\begin{align}
\mathcal{T}_{\mathrm s}(\bm{x},t)
&=
\mathcal{T}_{\mathrm s}^{\star}
\!\left(
\bm{x}+\bm{X}_{\mathrm p}(t)
\right),
\label{eq:Ts_loc_app}
\\
A_{\mathrm s}(\bm{x},t)
&=
1-\mathcal{T}_{\mathrm s}(\bm{x},t).
\label{eq:As_loc_app}
\end{align}

During secondary eclipse, the sky-plane intensity is
\begin{align}
I(\bm{x},t)
&=
I_{\mathrm s}(\bm{x},t)
+
\mathcal{T}_{\mathrm s}(\bm{x},t)\,I_{\mathrm p}(\bm{x})
\notag\\
&=
I_{\mathrm s}(\bm{x},t)
+
\left[1-A_{\mathrm s}(\bm{x},t)\right]I_{\mathrm p}(\bm{x}).
\end{align}
The time derivative of the secondary-eclipse flux is therefore
\begin{align}
\dot{F}_{\mathrm{sec}}(t)
&=
\frac{d}{dt}\iint I(\bm{x},t)\,dx\,dy
\notag\\
&=
-\frac{d}{dt}\iint A_{\mathrm s}(\bm{x},t)\,I_{\mathrm p}(\bm{x})\,dx\,dy
\notag\\
&=
-\iint
\frac{\partial}{\partial t}A_{\mathrm s}(\bm{x},t)\,
I_{\mathrm p}(\bm{x})\,dx\,dy,
\label{eq:F_sec_dot_app}
\end{align}
where we used that the total stellar flux $\iint I_{\mathrm{s}}(\bm{x},t)\,dx\,dy$ and the total planetary flux $\iint I_{\mathrm{p}}(\bm{x})\,dx\,dy$ are constant in time by assumption.

We now adopt the same local knife-edge approximation as in Section~\ref{sec:radon_limit_transit}.
For notational simplicity, we again use the subscripts $\mathrm{i}$ and $\mathrm{e}$ to denote ingress and egress, respectively, now for the secondary eclipse.
For $j\in\{\mathrm{i},\mathrm{e}\}$, let $t_j$ be the time at which the planet center crosses the stellar limb, so that $\bm{X}_{\mathrm p}(t_j)$ lies on the stellar limb.
Let $\theta_j$ be the angle of the inward normal to the tangent line of the stellar limb at $\bm{X}_{\mathrm p}(t_j)$, measured counterclockwise from the positive $X$-axis.
Let
\begin{equation}
\bm{n}_j \equiv (\cos\theta_j,\sin\theta_j)^{\mathrm T}
\end{equation}
be the corresponding inward unit normal.

Near ingress or egress of secondary eclipse, the stellar occultation map can then be approximated by
\begin{equation}
A_{\mathrm s,j}^{\star}(\bm{X})
=
H\!\left(
\bm{n}_j\cdot\left(\bm{X}-\bm{X}_{\mathrm{p}}(t_j)\right)
\right),
\label{eq:As_knife_app}
\end{equation}
where $H$ is the Heaviside step function.
Unlike in the transit case, the stellar surface-brightness profile does not enter the derivative expressions, so no assumption of uniform stellar brightness is required here.

Using the chain rule as in Equation~(\ref{eq:Is_del_t}), together with the distributional derivative of the Heaviside function, we obtain
\begin{equation}
\frac{\partial}{\partial t}A_{\mathrm s}(\bm{x},t)
=
\bm{n}_j\cdot\bm{v}(t)\,
\delta\!\left(
\bm{n}_j\cdot\left(\bm{x}+\bm{X}_{\mathrm p}(t)-\bm{X}_{\mathrm{p}}(t_j)\right)
\right).
\label{eq:dAsdt_j_app}
\end{equation}
Substituting this into Equation~(\ref{eq:F_sec_dot_app}) and comparing with the Radon-transform definition in Equation~(\ref{eq:Radon}), we obtain
\begin{equation}
\dot{F}_{\mathrm{sec},j}(t)
=
-\bm{n}_j\cdot\bm{v}(t)\,
\mathcal{R}[I_{\mathrm p}]
\!\left(
s_j(t),\,
\theta_j
\right),
\label{eq:Fdot_sec_Radon_j_app}
\end{equation}
where
\begin{equation}
s_j(t)
=
-\bm{n}_j\cdot\left(\bm{X}_{\mathrm{p}}(t)-\bm{X}_{\mathrm{p}}(t_j)\right).
\end{equation}

\begin{figure*}[ht!]
\epsscale{1.1}
\plotone{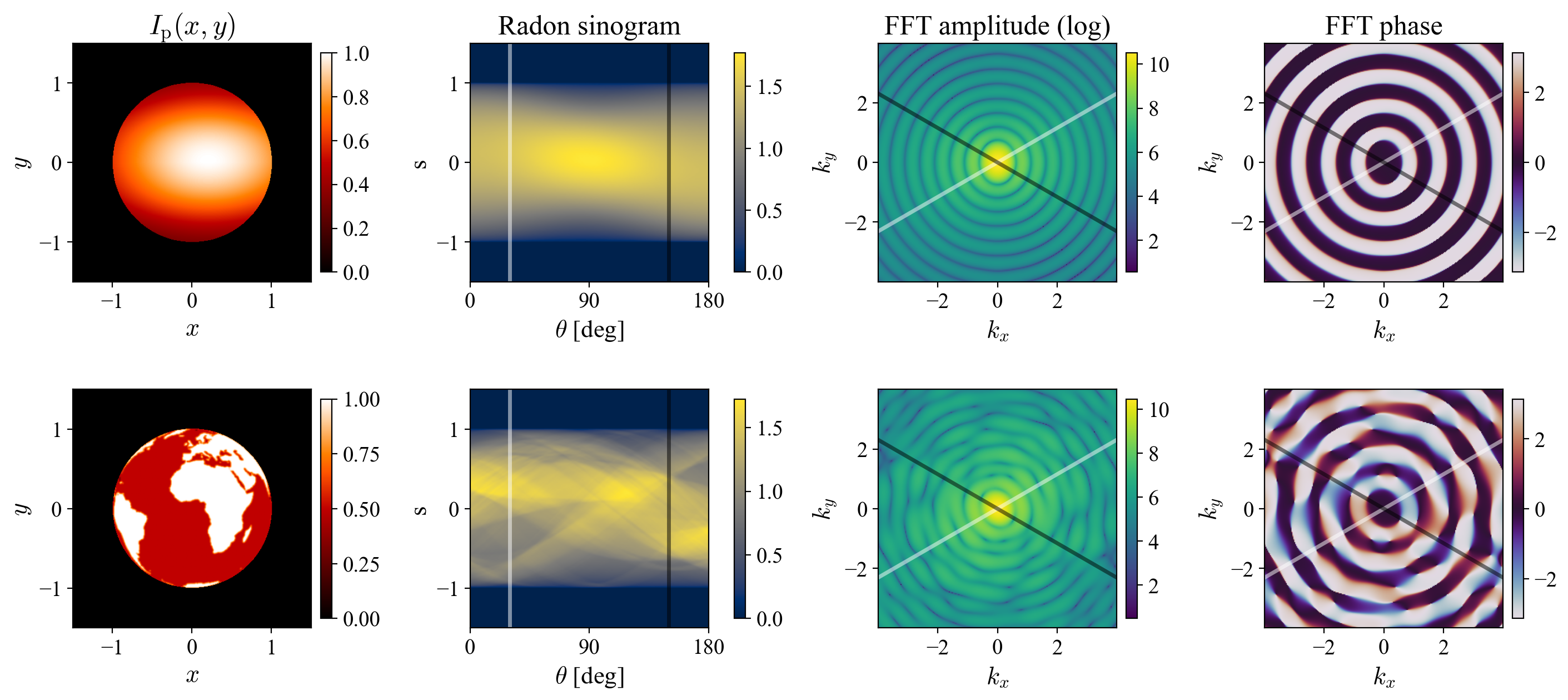}
\caption{
Radon projections and Fourier-domain representations of two synthetic spatial patterns of $I_{\mathrm{p}}(x,y)$.
As in Figure~\ref{fig:radon_fft}, each row shows the spatial map, its Radon sinogram, the log-scaled Fourier amplitude, and the Fourier phase.
The white and black lines indicate the Fourier slices accessible for an impact parameter of $b \sim 0.5$.
\label{fig:radon_fft_sec}}
\end{figure*}

Thus, in the Radon-transform limit, the time derivative of the secondary-eclipse light curve traces a Radon-transform projection of the planetary brightness map $I_{\mathrm p}(\bm{x})$.
Ingress and egress therefore provide at most two such projections.
Accordingly, the Fourier-domain null space has the same geometric structure as discussed in Section~\ref{sec:nullspace} and illustrated in Figure~\ref{fig:radon_fft_sec}.
In particular, any component of the brightness map whose Fourier transform vanishes on the observed slices is invisible in the idealized limit.

If we further approximate the projected velocity as constant over ingress or egress,
\begin{equation}
\bm{X}_{\mathrm{p}}(t)
\simeq
\bm{X}_{\mathrm{p}}(t_j)+\bm{v}(t_j)(t-t_j),
\end{equation}
then the temporal Fourier transform yields
\begin{equation}
\widetilde{\dot{F}_{\mathrm{sec},j}}(\nu)
=
-\sigma_j\,e^{-2\pi i\nu t_j}
\widehat{I_{\mathrm p}}
\!\left(
-\frac{\nu\cos\theta_j}{\alpha_j},
-\frac{\nu\sin\theta_j}{\alpha_j}
\right),
\label{eq:sec_Radon_FT_j_app}
\end{equation}
where
\begin{equation}
\alpha_j = \bm{n}_j\cdot\bm{v}(t_j),
\qquad
\sigma_j = \frac{\alpha_j}{|\alpha_j|}.
\end{equation}

A detailed discussion of eclipse-mapping null spaces in a spherical-harmonic framework is given by \citet{Challener_2023}.
Although their formulation includes planetary rotation and phase-curve information, the qualitative picture is broadly consistent with the one derived here.
\citet{2024MNRAS.528..596B} adopted the same straight-edge approximation to develop an analytical theory for the observable spatial modes and the spatial resolution attainable with eclipse mapping.

\section{Approximate Limb Darkening as a Convolution}
\setcounter{equation}{0}
\label{app:limb_darkening}

In this Appendix, we derive the limb-darkening expression quoted in Section~\ref{sec:limb_darkening}.
Under the same local knife-edge approximation adopted in Section~\ref{sec:radon_limit_transit}, stellar limb darkening does not introduce new projection angles.
Instead, it replaces the ideal Radon-transform relation by a one-dimensional weighting, or equivalently a convolution, along the Radon-offset direction.

We again use the subscripts $\mathrm{i}$ and $\mathrm{e}$ to denote ingress and egress, respectively.
For $j\in\{\mathrm{i},\mathrm{e}\}$, let $t_j$ be the time at which the planet center crosses the stellar limb, so that $\bm{X}_{\mathrm p}(t_j)$ lies on the stellar limb.
Let $\theta_j$ be the angle of the inward normal to the tangent line of the stellar limb at $\bm{X}_{\mathrm p}(t_j)$, measured counterclockwise from the positive $X$-axis.
Let
\begin{equation}
\bm{n}_j \equiv (\cos\theta_j,\sin\theta_j)^{\mathrm T}
\end{equation}
be the corresponding inward unit normal.

We introduce a local one-dimensional brightness profile along the inward normal direction.
Specifically, we define
\begin{equation}
u_j(\bm{X})
\equiv
\bm{n}_j\cdot\left(\bm{X}-\bm{X}_{\mathrm p}(t_j)\right),
\label{eq:u_ld_j}
\end{equation}
and approximate the stellar intensity near ingress or egress by
\begin{equation}
I_{\mathrm s}^{\star}(\bm{X})
\simeq
\Psi_j\!\left(u_j(\bm{X})\right),
\label{eq:Is_ld_local_general}
\end{equation}
where $\Psi_j(u)$ is the local stellar brightness profile as a function of distance from the locally linearized limb.
In the uniform-brightness limit, $\Psi_j(u)=I_0 H(u)$.

Expressed in planet-centered coordinates, Equation~(\ref{eq:Is_loc}) becomes
\begin{equation}
I_{\mathrm s}(\bm{x},t)
\simeq
\Psi_j\!\left(\bm{n}_j\cdot\bm{x}-s_j(t)\right),
\label{eq:Is_ld_planet_general}
\end{equation}
where
\begin{equation}
s_j(t)
\equiv
-\bm{n}_j\cdot\left(\bm{X}_{\mathrm p}(t)-\bm{X}_{\mathrm p}(t_j)\right)
\label{eq:sj_ld_def}
\end{equation}
is the same Radon offset introduced in Equation~(\ref{eq:sj_def}).
Applying the chain rule to Equation~(\ref{eq:Is_ld_planet_general}) gives
\begin{equation}
\frac{\partial}{\partial t}I_{\mathrm s}(\bm{x},t)
=
\bm{n}_j\cdot\bm{v}(t)\,
\Psi_j'\!\left(\bm{n}_j\cdot\bm{x}-s_j(t)\right),
\label{eq:dIsdt_ld_general}
\end{equation}
where $\Psi_j'(u)\equiv d\Psi_j/du$.

Substituting Equation~(\ref{eq:dIsdt_ld_general}) into Equation~(\ref{eq:F_tr_dot}), we obtain
\begin{equation}
\dot{F}_j(t)
=
-\bm{n}_j\cdot\bm{v}(t)
\iint
A_{\mathrm p}(\bm{x})\,
\Psi_j'\!\left(\bm{n}_j\cdot\bm{x}-s_j(t)\right)
\,dx\,dy.
\label{eq:Fdot_ld_general_pre}
\end{equation}
Using
\begin{equation}
\Psi_j'(u)
=
\int_{\mathbb{R}}
\Psi_j'(p)\,\delta(u-p)\,dp,
\end{equation}
this becomes
\begin{align}
\dot{F}_j(t)
&=
-\bm{n}_j\cdot\bm{v}(t)
\iiint
A_{\mathrm p}(\bm{x})\,
\Psi_j'(p)\,
\delta\!\left(\bm{n}_j\cdot\bm{x}-s_j(t)-p\right)
\,dp\,dx\,dy
\notag\\
&=
-\bm{n}_j\cdot\bm{v}(t)
\int_{\mathbb{R}}
\Psi_j'(p)\,
\mathcal{R}[A_{\mathrm p}]\!\left(s_j(t)+p,\theta_j\right)
\,dp.
\label{eq:Fdot_ld_general}
\end{align}
Thus, limb darkening does not change the projection angle $\theta_j$.
Instead, it mixes nearby Radon offsets along the same projection direction.
Therefore, in the local knife-edge approximation, stellar limb darkening acts as a one-dimensional convolution kernel along the Radon-offset direction.

To obtain a convolution directly in time, we make one additional approximation: over ingress or egress, the projected velocity is approximately constant.
We then write
\begin{equation}
\bm{X}_{\mathrm p}(t)
\simeq
\bm{X}_{\mathrm p}(t_j)+\bm{v}(t_j)(t-t_j),
\end{equation}
and define
\begin{equation}
\alpha_j\equiv\bm{n}_j\cdot\bm{v}(t_j),
\end{equation}
so that
\begin{equation}
s_j(t)\simeq -\alpha_j(t-t_j).
\label{eq:sj_linear_ld}
\end{equation}
Substituting Equation~(\ref{eq:sj_linear_ld}) into Equation~(\ref{eq:Fdot_ld_general}) gives
\begin{equation}
\dot{F}_j(t)
\simeq
-\alpha_j
\int_{\mathbb{R}}
\Psi_j'(p)\,
\mathcal{R}[A_{\mathrm p}]\!\left(-\alpha_j(t-t_j)+p,\theta_j\right)
\,dp.
\label{eq:Fdot_ld_time_pre}
\end{equation}
Defining
\begin{equation}
q\equiv \frac{p}{\alpha_j},
\end{equation}
and
\begin{equation}
K_j(q)\equiv |\alpha_j|\,\Psi_j'(\alpha_j q),
\label{eq:Kj_def}
\end{equation}
we obtain
\begin{equation}
\dot{F}_j(t)
\simeq
-\alpha_j
\int_{\mathbb{R}}
K_j(q)\,
\mathcal{R}[A_{\mathrm p}]\!\left(-\alpha_j(t-t_j-q),\theta_j\right)
\,dq.
\label{eq:Fdot_ld_time_conv}
\end{equation}
Thus, after the additional constant-velocity approximation, limb darkening appears as a one-dimensional convolution in time.

Taking the temporal Fourier transform of Equation~(\ref{eq:Fdot_ld_time_conv}) yields
\begin{align}
\widetilde{\dot{F}_j}(\nu)
&=
-\sigma_j\,
e^{-2\pi i\nu t_j}\,
\widetilde{K_j}(\nu)\,
\widehat{A_{\mathrm p}}
\!\left(
-\frac{\nu\cos\theta_j}{\alpha_j},
-\frac{\nu\sin\theta_j}{\alpha_j}
\right),
\label{eq:Fdot_limbdark_FT_j}
\end{align}
where
\begin{equation}
\sigma_j \equiv \frac{\alpha_j}{|\alpha_j|}.
\end{equation}
Therefore, limb darkening does not change the locations of the accessible Fourier slices.
It only multiplies those slices by a one-dimensional transfer function determined by the local limb-darkening profile.

\section{Corrections from Stellar-Limb Curvature}
\setcounter{equation}{0}
\label{app:curvature_corrections}

In this Appendix, we derive the leading correction to the Radon-transform limit when the stellar limb is no longer approximated as locally straight.
We retain uniform stellar surface brightness, so the only correction considered here comes from the curvature of the stellar limb.

We again use the subscripts $\mathrm{i}$ and $\mathrm{e}$ to denote ingress and egress, respectively.
For $j\in\{\mathrm{i},\mathrm{e}\}$, let $\theta_j$ be the angle of the inward normal to the stellar limb at the reference point corresponding to time $t_j$, and define
\begin{equation}
\bm{n}_j\equiv (\cos\theta_j,\sin\theta_j)^{\mathrm T},
\qquad
\bm{\tau}_j\equiv (-\sin\theta_j,\cos\theta_j)^{\mathrm T},
\end{equation}
so that $\bm{\tau}_j$ is the corresponding unit tangent vector.
We also denote by $\kappa_j$ the local curvature of the stellar limb at that point.
For a spherical star, $\kappa_j=1/R_{\mathrm{s}}$.

Let
\begin{equation}
\xi \equiv \bm{n}_j\cdot\left(\bm{X}-\bm{X}_{\mathrm p}(t_j)\right),
\qquad
\eta \equiv \bm{\tau}_j\cdot\left(\bm{X}-\bm{X}_{\mathrm p}(t_j)\right)
\end{equation}
be local normal and tangential coordinates centered on the reference point.
To second order in $\eta$, the stellar boundary is approximated by
\begin{equation}
\xi = \frac{\kappa_j}{2}\eta^2.
\label{eq:boundary_parabola_app}
\end{equation}
Thus, for uniform stellar brightness, the stellar intensity can be approximated locally as
\begin{equation}
I_{\mathrm s}^{\star}(\bm{X}) \simeq I_0\,H\!\left(\xi-\frac{\kappa_j}{2}\eta^2\right).
\label{eq:Is_curved_local_app}
\end{equation}
Equation~(\ref{eq:Is_curved_local_app}) reduces to the knife-edge approximation when $\kappa_j\to 0$.

Now express the stellar intensity in planet-centered coordinates.
Define the planet-center displacement relative to the reference point by
\begin{equation}
\Delta\bm{X}_j(t)\equiv \bm{X}_{\mathrm p}(t)-\bm{X}_{\mathrm p}(t_j),
\end{equation}
and decompose it as
\begin{equation}
\xi_j(t)\equiv \bm{n}_j\cdot \Delta\bm{X}_j(t),
\qquad
\eta_j(t)\equiv \bm{\tau}_j\cdot \Delta\bm{X}_j(t).
\label{eq:xi_eta_def_app}
\end{equation}
Note that $\xi_j(t)$ is simply the negative of the Radon offset introduced in the main text:
\begin{equation}
\xi_j(t)=-s_j(t),
\end{equation}
where $s_j(t)$ is defined in Equation~(\ref{eq:sj_def}).

Then
\begin{equation}
I_{\mathrm s}(\bm{x},t) \simeq I_0\,H\!\left(\bm{n}_j\cdot\bm{x}+\xi_j(t)-\frac{\kappa_j}{2}\bigl(\bm{\tau}_j\cdot\bm{x}+\eta_j(t)\bigr)^2\right).
\label{eq:Is_curved_planet_app}
\end{equation}
Differentiating Equation~(\ref{eq:Is_curved_planet_app}) with respect to time, and using the distributional derivative of the Heaviside function, gives
\begin{equation}
\frac{\partial}{\partial t}I_{\mathrm s}(\bm{x},t)=I_0\,\delta\!\left(\bm{n}_j\cdot\bm{x}+\xi_j(t)-\frac{\kappa_j}{2}\bigl(\bm{\tau}_j\cdot\bm{x}+\eta_j(t)\bigr)^2\right)\left[\dot{\xi}_j(t)-\kappa_j\dot{\eta}_j(t)\bigl(\bm{\tau}_j\cdot\bm{x}+\eta_j(t)\bigr)\right].
\label{eq:dIsdt_curved_app}
\end{equation}
Substituting this into Equation~(\ref{eq:F_tr_dot}) yields
\begin{equation}
\dot{F}_j(t)=-I_0\iint A_{\mathrm p}(\bm{x})\Bigl[\dot{\xi}_j(t)-\kappa_j\dot{\eta}_j(t)\bigl(\bm{\tau}_j\cdot\bm{x}+\eta_j(t)\bigr)\Bigr]\delta\!\left(\bm{n}_j\cdot\bm{x}+\xi_j(t)-\frac{\kappa_j}{2}\bigl(\bm{\tau}_j\cdot\bm{x}+\eta_j(t)\bigr)^2\right)\,dx\,dy.
\label{eq:Fdot_curved_app}
\end{equation}
Expanding the argument gives
\begin{align}
\bm{n}_j\cdot\bm{x}+\xi_j(t)-\frac{\kappa_j}{2}\bigl(\bm{\tau}_j\cdot\bm{x}+\eta_j(t)\bigr)^2
&=\bm{n}_j\cdot\bm{x}+\xi_j(t)-\frac{\kappa_j}{2}\eta_j(t)^2-\kappa_j\eta_j(t)\,\bm{\tau}_j\cdot\bm{x}-\frac{\kappa_j}{2}\bigl(\bm{\tau}_j\cdot\bm{x}\bigr)^2 \notag \\
&=\bm{n}_j\cdot\bm{x}-s_j^{\mathrm{eff}}(t)-\kappa_j\eta_j(t)\,\bm{\tau}_j\cdot\bm{x}-\frac{\kappa_j}{2}\bigl(\bm{\tau}_j\cdot\bm{x}\bigr)^2,
\label{eq:Is_curved_eff_app}
\end{align}
where we absorb the constant term into an effective offset,
\begin{equation}
s_j^{\mathrm{eff}}(t)\equiv -\xi_j(t)+\frac{\kappa_j}{2}\eta_j(t)^2.
\label{eq:s_eff_app}
\end{equation}

We now treat the tangentially dependent terms as perturbations around the ideal slice at angle $\theta_j$ and offset $s_j^{\mathrm{eff}}(t)$.
Using, in the distributional sense,
\begin{equation}
\delta(a-\varepsilon b)=\delta(a)-\varepsilon b\,\delta'(a)+\mathcal O(\varepsilon^2),
\end{equation}
with
\begin{equation}
a=\bm{n}_j\cdot\bm{x}-s_j^{\mathrm{eff}}(t),
\qquad
\varepsilon b=\kappa_j\eta_j(t)\,\bm{\tau}_j\cdot\bm{x}+\frac{\kappa_j}{2}\bigl(\bm{\tau}_j\cdot\bm{x}\bigr)^2,
\end{equation}
and retaining terms through first order in $\kappa_j$, we obtain
\begin{align}
\dot{F}_j(t)
=&-I_0\Bigl[\dot{\xi}_j(t)-\kappa_j\eta_j(t)\dot{\eta}_j(t)\Bigr]\iint A_{\mathrm p}(\bm{x})\,\delta\!\left(\bm{n}_j\cdot\bm{x}-s_j^{\mathrm{eff}}(t)\right)\,dx\,dy \notag \\
&+I_0\kappa_j\dot{\eta}_j(t)\iint \bigl(\bm{\tau}_j\cdot\bm{x}\bigr)A_{\mathrm p}(\bm{x})\,\delta\!\left(\bm{n}_j\cdot\bm{x}-s_j^{\mathrm{eff}}(t)\right)\,dx\,dy \notag \\
&+I_0\kappa_j\dot{\xi}_j(t)\iint \left[\eta_j(t)\,\bm{\tau}_j\cdot\bm{x}+\frac{1}{2}\bigl(\bm{\tau}_j\cdot\bm{x}\bigr)^2\right]A_{\mathrm p}(\bm{x})\,\delta'\!\left(\bm{n}_j\cdot\bm{x}-s_j^{\mathrm{eff}}(t)\right)\,dx\,dy+\mathcal O(\kappa_j^2).
\label{eq:Fdot_curved_expand_app}
\end{align}

To express these terms compactly, define
\begin{equation}
M_{m}[f](s,\theta)\equiv \iint \bigl(\bm{\tau}(\theta)\cdot\bm{x}\bigr)^m f(\bm{x})\,\delta\!\left(\bm{n}(\theta)\cdot\bm{x}-s\right)\,dx\,dy,
\label{eq:Mm_def_app}
\end{equation}
where
\begin{equation}
\bm{n}(\theta)\equiv (\cos\theta,\sin\theta)^{\mathrm T},
\qquad
\bm{\tau}(\theta)\equiv (-\sin\theta,\cos\theta)^{\mathrm T}.
\end{equation}
The case $m=0$ is simply the Radon transform:
\begin{equation}
M_{0}[f](s,\theta)=\mathcal R[f](s,\theta).
\end{equation}
Using
\begin{equation}
\iint g(\bm{x})\,\delta'\!\left(\bm{n}_j\cdot\bm{x}-s\right)\,dx\,dy=-\frac{\partial}{\partial s}\left[\iint g(\bm{x})\,\delta\!\left(\bm{n}_j\cdot\bm{x}-s\right)\,dx\,dy\right],
\end{equation}
Equation~(\ref{eq:Fdot_curved_expand_app}) becomes
\begin{align}
\dot{F}_j(t)=&-I_0\Bigl[\dot{\xi}_j(t)-\kappa_j\eta_j(t)\dot{\eta}_j(t)\Bigr]M_{0}[A_{\mathrm p}]\!\left(s_j^{\mathrm{eff}}(t), \theta_j\right) \notag \\
&+I_0\kappa_j\dot{\eta}_j(t)\,M_{1}[A_{\mathrm p}]\!\left(s_j^{\mathrm{eff}}(t), \theta_j\right) \notag \\
&-I_0\kappa_j\dot{\xi}_j(t)\,\frac{\partial}{\partial s}\left[\eta_j(t)\,M_{1}[A_{\mathrm p}](s, \theta_j)+\frac{1}{2}M_{2}[A_{\mathrm p}](s, \theta_j)\right]_{s=s_j^{\mathrm{eff}}(t)}+\mathcal O(\kappa_j^2).
\label{eq:Fdot_curved_moment_app}
\end{align}

Equation~(\ref{eq:Fdot_curved_moment_app}) is the desired first-order correction to the ideal Radon-transform limit.
The first term is the usual Radon-transform contribution, evaluated at the curvature-corrected offset.
The second term introduces sensitivity to the first tangential moment of the map along the ideal slice, while the third term couples neighboring offsets through the derivative of a combination of the first and second tangential moments.

The Fourier-space meaning of these quantities is also straightforward.
Define the one-dimensional Fourier transform of $M_m[f](s,\theta)$ by
\begin{equation}
\widetilde{M}_{m}[f](k, \theta)\equiv \int_{\mathbb R} M_{m}[f](s, \theta)\,e^{-2\pi i k s}\,ds.
\end{equation}
Then
\begin{equation}
\widetilde{M}_{m}[f](k, \theta_j)=\iint \bigl(\bm{\tau}_j\cdot\bm{x}\bigr)^m f(\bm{x})\,e^{-2\pi i k\,\bm{n}_j\cdot\bm{x}}\,dx\,dy.
\end{equation}
If we introduce Fourier coordinates $(k_{\parallel},k_{\perp})$ aligned with $(\bm{n}_j,\bm{\tau}_j)$, so that
\begin{equation}
\widehat{f}(k_{\parallel},k_{\perp})\equiv \iint f(\bm{x})\,e^{-2\pi i (k_{\parallel}\bm{n}_j+k_{\perp}\bm{\tau}_j)\cdot\bm{x}}\,dx\,dy,
\end{equation}
then multiplication by $\bm{\tau}_j\cdot\bm{x}$ in real space corresponds to differentiation with respect to $k_{\perp}$ in Fourier space, and hence
\begin{equation}
\widetilde{M}_{m}[f](k, \theta_j)=\left(\frac{i}{2\pi}\right)^m \frac{\partial^m}{\partial k_{\perp}^m}\widehat{f}(k_{\parallel},k_{\perp})\Big|_{k_{\parallel}=k,\;k_{\perp}=0}.
\label{eq:Mm_fourier_app}
\end{equation}
Thus, $M_{0}$ gives the value of $\widehat{f}$ on the ideal radial slice, $M_{1}$ gives the first transverse derivative of $\widehat{f}$ about that slice, and $M_{2}$ gives the second transverse derivative.
The derivative with respect to $s$ in Equation~(\ref{eq:Fdot_curved_moment_app}) corresponds, after Fourier transformation, to multiplication by $2\pi i k_{\parallel}$.

Therefore, small stellar-limb curvature does not provide entirely new angular coverage in the Fourier plane.
Instead, it makes the light curve weakly sensitive to derivatives of $\widehat{A}_{\mathrm p}$ transverse to the ideal slices.
Some map perturbations that are invisible in the ideal Radon-transform limit therefore produce perturbatively small curvature-induced responses, while Fourier-space structure far from the measured slices remains only weakly constrained.
Within this perturbative picture, the overlap between the information carried by the ingress and egress measurements is minimized when the two ideal slices are orthogonal, which occurs at $b \sim 1/\sqrt{2}$.
In that case, the region of Fourier space that becomes weakly accessible through curvature effects is broadened in the most complementary way.

\bibliography{sample701}{}
\bibliographystyle{aasjournalv7}

\end{document}